\documentclass[preprint]{aastex63}

\makeatletter
\newcommand\DefineObj[3][\@empty]{%
  \expandafter\newcommand\csname pkgwobj@l#2\endcsname{#3}%
  \ifx\@empty#1%
    \expandafter\newcommand\csname pkgwobj@s#2\endcsname{#3}%
  \else%
    \expandafter\newcommand\csname pkgwobj@s#2\endcsname{#1}%
  \fi}%
\newcommand{\pkgw@printobj@long}[1]{%
  \expandafter\ifx\csname pkgwobj@l#1\endcsname\relax%
    \textbf{[unknown object!]}%
  \else%
    \csname pkgwobj@l#1\endcsname%
  \fi}%
\newcommand{\pkgw@printobj@short}[1]{%
  \expandafter\ifx\csname pkgwobj@l#1\endcsname\relax%
    \textbf{[unknown object!]}%
  \else%
    \csname pkgwobj@s#1\endcsname%
  \fi}%
\newcommand{\obj}{\@ifstar{\pkgw@printobj@long}{\pkgw@printobj@short}}%
\makeatother
\DefineObj[2MASS~J0045$+$16]{2m0045}{2MASS~J00452143$+$1634446}
\DefineObj[2MASS~J0501$-$00]{2m0501}{2MASS~J05012406$-$0010452}
\DefineObj[2MASS~J1425$-$36]{2m1425}{2MASS~J14252798$-$3650229}
\usepackage{gensymb}

\graphicspath{{./}{}}

\shorttitle{Young L Variables with Spitzer}
\shortauthors{Vos et al.}

\begin{document}

\title{\textit{Spitzer} Variability Properties of Low-Gravity L Dwarfs}

\correspondingauthor{Johanna M. Vos}
\email{jvos@amnh.org}

\author[0000-0003-0489-1528]{Johanna M. Vos}
\affil{Department of Astrophysics, American Museum of Natural History, Central Park West at 79th Street, New York, NY 10024, USA}

\author[0000-0003-4614-7035]{Beth A. Biller}
\affiliation{SUPA, Institute for Astronomy, University of Edinburgh, Blackford Hill, Edinburgh EH9 3HJ, UK}
\affiliation{Centre for Exoplanet Science, University of Edinburgh, Edinburgh, UK}

\author{Katelyn N. Allers}
\affiliation{Department of Physics and Astronomy, Bucknell University, Lewisburg, PA 17837, USA}

\author[0000-0001-6251-0573]{Jacqueline K. Faherty}
\affiliation{Department of Astrophysics, American Museum of Natural History, Central Park West at 79th Street, New York, NY 10024, USA}

\author[0000-0003-2232-7664]{Michael C. Liu}
\affiliation{Institute for Astronomy, University of Hawaii at Manoa, Honolulu, HI 96822, USA}

\author[0000-0003-3050-8203]{Stanimir Metchev}
\affiliation{Department of Physics \& Astronomy and Centre for Planetary Science and Exploration, The University of Western Ontario, London, Ontario N6A 3K7, Canada}

\author[0000-0001-6377-8272]{Simon Eriksson}
\affiliation{Department of Astronomy, Stockholm University, AlbaNova University Center, SE-106 91 Stockholm, Sweden}

\author[0000-0003-0192-6887]{Elena Manjavacas}
\affiliation{W.M. Keck Observatory, 65-1120 Mamalahoa Highway, Kamuela, HI 96743, USA}

\author[0000-0001-9823-1445]{Trent J. Dupuy}
\affiliation{Gemini Observatory, Northern Operations Center, 670 N. A'ohoku Place, Hilo, HI 96720 USA}

\author[0000-0001-8345-593X]{Markus Janson}
\affiliation{Department of Astronomy, Stockholm University, AlbaNova University Center, SE-106 91 Stockholm, Sweden}

\author{Jacqueline Radigan-Hoffman}
\affiliation{Utah Valley University, 800 West University Parkway, Orem, UT 84058, USA}

\author{Ian Crossfield}
\affiliation{Physics \& Astronomy Department, University of Kansas, Lawrence, KS, USA}

\author{Micka\"el Bonnefoy}
\affiliation{Univ. Grenoble Alpes, IPAG; CNRS, IPAG, 38000 Grenoble, France}

\author[0000-0003-0562-1511]{William M. J. Best}
\affiliation{The University of Texas at Austin, Department of Astronomy, 2515 Speedway C1400, Austin, TX 78712, USA}

\author[0000-0002-8546-9128]{Derek Homeier}
\affiliation{Zentrum f\"ur Astronomie der Universit\"at Heidelberg, Landessternwarte, K\"onigstuhl 12, D-69117 Heidelberg, Germany}

\author[0000-0001-5347-7062]{Joshua E. Schlieder}
\affiliation{Exoplanets and Stellar Astrophysics Laboratory, Code 667, NASA Goddard Space Flight Center, Greenbelt, MD 20771, USA}

\author{Wolfgang Brandner}
\affiliation{Max-Planck-Institut f\"ur Astronomie, K\"onigstuhl 17, D-69117 Heidelberg, Germany}

\author{Thomas Henning}
\affiliation{Max-Planck-Institut f\"ur Astronomie, K\"onigstuhl 17, D-69117 Heidelberg, Germany}

\author[0000-0002-7520-8389]{Mariangela Bonavita}
\affiliation{SUPA, Institute for Astronomy, University of Edinburgh, Blackford Hill, Edinburgh EH9 3HJ, UK}
\affiliation{Centre for Exoplanet Science, University of Edinburgh, Edinburgh, UK}

\author[0000-0003-3306-1486]{Esther Buenzli}
\affiliation{Institute for Particle Physics and Astrophysics, ETH Zurich, Wolfgang-Pauli- Strasse 27, 8093, Zurich, Switzerland}

\begin{abstract}
We present \textit{Spitzer Space Telescope} variability monitoring observations of three low-gravity L dwarfs with previous detections of variability in the near-IR, \obj{2m0045}, \obj{2m0501} and \obj{2m1425}. We detect significant, periodic variability in two of our targets, \obj{2m0045} and \obj{2m0501}. We do not detect variability in \obj{2m1425}. 
Combining our new rotation periods with rotational velocities, we calculate inclination angles of $22\pm1^{\circ}$, ${60^{+13 }_{-8}} ^{\circ}$ and $52^{+19}_{-13}~^{\circ}$ for  \obj{2m0045}, \obj{2m0501} and \obj{2m1425} respectively. Our three new objects are consistent with the tentative relations between inclination, amplitude and color anomaly previously reported. Objects with the highest variability amplitudes are inclined equator-on, while the maximum observed amplitude decreases as the inclination angle decreases. We also find a correlation between the inclination angle and $(J-K)_{\mathrm{2MASS}}$ color anomaly for the sample of objects with measured inclinations. 
Compiling the entire sample of brown dwarfs with \textit{Spitzer} variability detections, we find no enhancement in amplitude for young, early-L dwarfs compared to the field dwarf population. We find a possible enhancement in amplitude of low-gravity late-L dwarfs at $4.5~\mu$m. We do not find a correlation between amplitude ratio and spectral type for field dwarfs or for the young population. 
Finally, we compile the rotation periods of a large sample of brown dwarfs with ages 1 Myr to 1 Gyr and compare the rotation rates predicted by evolutionary models assuming angular momentum conservation. We find that the rotation rates of the current sample of brown dwarfs fall within the expected range set by evolutionary models and breakup limits.

\end{abstract}

\keywords{brown dwarfs -- stars: rotation -- stars: variables: general -- techniques: photometric}

\section{Introduction} \label{sec:intro}

Photometric variability monitoring directly probes atmospheric features in exoplanet and brown dwarf atmospheres, as it is sensitive to the spatial distribution of inhomogeneities such as condensate clouds, compositional and/or temperature fluctuations \citep{Marley2012, Tremblin2016} as a planet rotates. {Variability studies of field brown dwarfs have begun to reveal the complex, evolving nature of their atmospheres \citep[][and references therein]{Biller2017, Artigau2018}. Recent variability monitoring with \textit{Spitzer} and the \textit{Very Large Array} has provided the first direct measurements of the wind speed on a brown dwarf, finding a strong eastward wind of $\sim650$m/s for the cool T dwarf 2MASS J10475385+2124234 \citep{Allers2020}.} 
For the majority of directly-imaged planets however, the contrast between host star and planet make it difficult to obtain sufficiently high S/N photometry to allow precise variability monitoring \citep{Apai2016}.
Young, free-floating brown dwarfs provide an excellent analog to directly-imaged planets. 
They share remarkably similar masses, radii and spectra with the directly-imaged planets \citep{Faherty2016, Liu2013, Liu2016} and can be observed in detail with current facilities.

In recent years variability studies of the low-gravity exoplanet analogs have revealed important insights into their atmospheres. \citet{Metchev2015a} noted a tentative correlation between low-gravity and high-amplitude variability in a sample of six mid-L type brown dwarfs as part of a larger \textit{Spitzer} variability survey. A number of high-amplitude variability detections have since been reported in free-floating planetary-mass objects \citep{Biller2015, Lew2016, Vos2018, Schneider2018} and wide, low-gravity companions \citep{Zhou2016, Zhou2019, Zhou2020, Bowler2020}. 
In \citet{Vos2019} we carried out a large ground-based, $J$-band survey for variability in isolated low-mass brown dwarfs, finding a $30\%$ variability fraction among low-gravity L dwarfs, compared to $11\%$ for the higher gravity, field brown dwarf population surveyed by \citet{Radigan2014}. This may be a result of the high altitude clouds found in low-gravity atmospheres \citep{Marley2012} providing a higher contrast ratio between cloud layers which would enhance the amplitude produced by inhomogeneities in the cloud deck.

As part of a large survey for $J$-band variability in low-mass brown dwarfs with NTT/SofI, we detected variability in 6 young L-type dwarfs \citep{Vos2019}. We have carried out follow-up monitoring of 2 detections, the late-L objects PSO J318.5338−22.8603 (PSO 318.5-22) and 2MASS J2244316+204343 (2M2244+20)  in previous \textit{Spitzer} cycles \citep{Biller2018, Vos2018} and here we present follow-up mid-IR monitoring of the three L2--L5 spectral type detections from this survey -- \obj*{2m0045} (hereafter \obj{2m0045}), \obj*{2m0501} (hereafter \obj{2m0501}) and \obj*{2m1425} (hereafter \obj{2m1425}).

All three objects show robust evidence of youth and/or low-gravity. 
The L2 object \obj{2m0045} is a confirmed member of the $50~$Myr old Argus association \citep{Faherty2016, Liu2016}, and shows signs of very low-gravity in its spectrum \citep{Gagne2015c, Allers2013}. \citet{Riedel2019} find that the kinematics of \obj{2m0045} may be consistent with the younger $23~$Myr $\beta$ Pictoris moving group using their LACEwING group membership code \citep{Riedel2017}.
The L3 object \obj{2m0501} is not associated with a known young moving group but shows signs of very low gravity in its optical and IR spectra \citep{Cruz2009, Allers2013}. Both \citet{Liu2016} and \citet{Faherty2016} classify \obj{2m0501} as a young field object.
The L4 object \obj{2m1425} is a member of the $110-150~$Myr old AB Doradus moving group \citep{Faherty2016, Liu2016} and has been classified as an intermediate-gravity object based on its IR spectrum \citep{Gagne2015c}. All three targets have estimated evolutionary model masses of $20-25~M_{\mathrm{Jup}}$ based on their bolometric luminosities \citep{Faherty2016}. 

In addition, we obtained Gemini/GNIRS high-resolution spectra of \obj{2m0501} and re-reduced Keck/NIRSPEC spectra for \obj{2m0045} and \obj{2m0501}. In Section \ref{sec:rotvel} we describe the analysis of our high-resolution spectra, in Section \ref{sec:obs_datareduc} we describe the \textit{Spitzer} observations and data reduction, in Section \ref{sec:var_id} we describe how we identified variables and in Sections  \ref{sec:amplitudes}, \ref{sec:inclination} and \ref{sec:periods} we describe our results in the context of brown dwarf variability amplitudes, inclination angles and rotation periods.

\section{Target Rotational Velocities and Maximum Periods}
\label{sec:rotvel}
\obj{2m0045} and \obj{2m1425} have high-resolution NIRSPEC-7 data from Keck/NIRSPEC in the Keck Observatory Archive (Program IDs C34NS, N58NS; PI: Charbonneau).  We additionally obtained a high-dispersion spectrum of \obj{2m0501} using Gemini/GNIRS (PID: GN-2017B-Q-58). We used the 111 l/mm grating and the long camera ($0.05''$ per pixel) with a $0.10''$ slit, centered at $2.3~\mu$m. We obtained four 600 s exposures taken in an ABBA nod pattern. The data were reduced using a modified version of the {\sc redspec} package as described in \citet{Vos2017}. 

\begin{figure}[tb]
   \centering
   \includegraphics[scale=0.75]{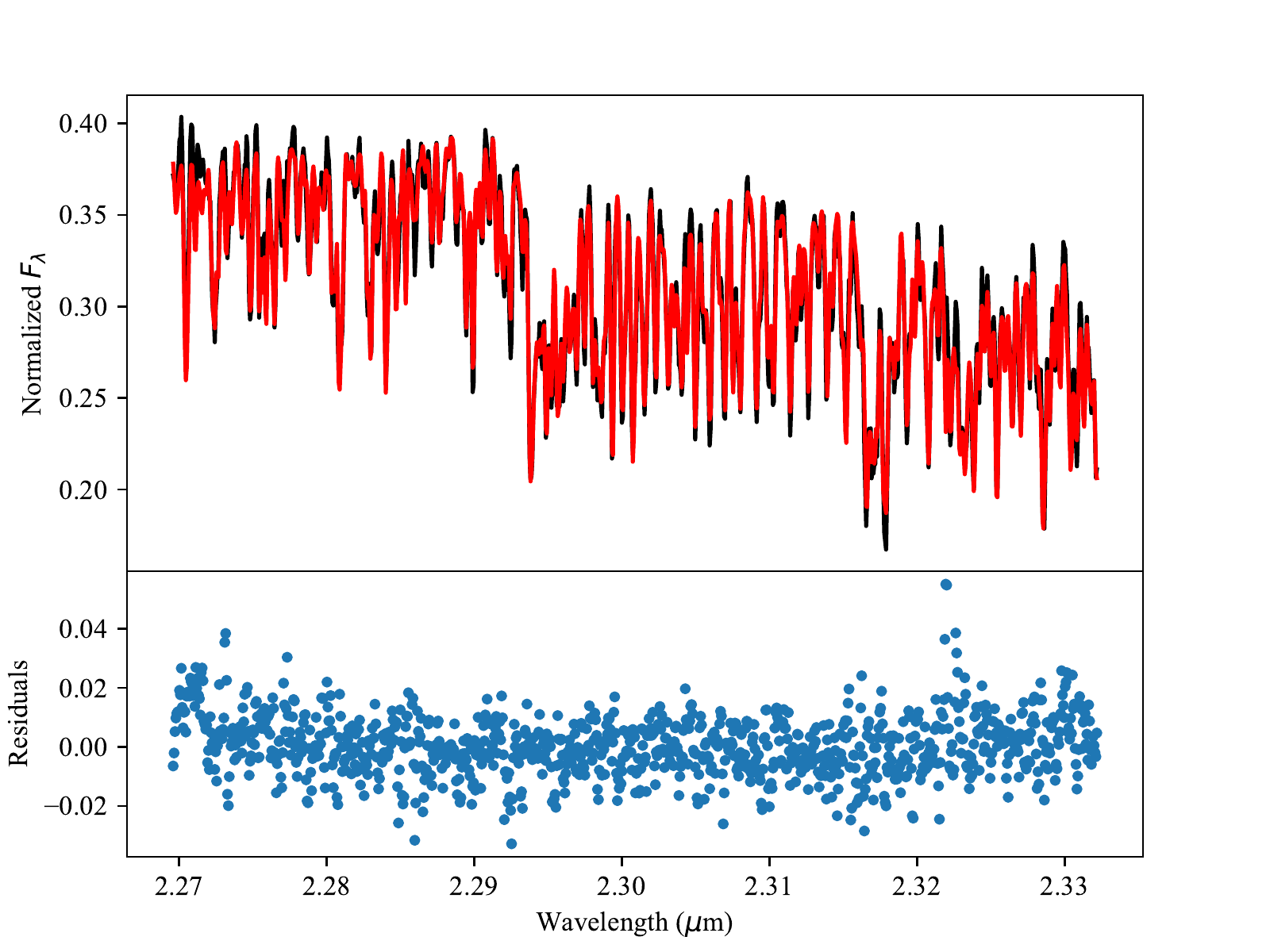}
   \caption{Top panel: Gemini/GNIRS spectrum of \obj{2m0501} shown in black, with our best fit model overplotted in red. Bottom panel: Residuals of the fit.}
   \label{fig:2M0501_vsini}
\end{figure}

We use the method outlined in our previous work \citep{Allers2016,Vos2017} to calculate the radial and rotational velocity of our three targets. We use forward modelling to simultaneously fit the wavelength solution, the scaling of telluric line depths, the FWHM of the instrument line spread function and the target radial and rotation velocity. The BT-Settl model atmospheres \citep{Allard2012} are used as the intrinsic spectrum for each target. We use MCMC methods to determine the posterior distributions for our forward model parameters. {To ensure that the median absolute residual of the fit agrees with the median uncertainty of our spectra, we include a systematic uncertainty in our analysis. We obtain a systematic uncertainty of $1.9\%$ for \obj{2m0045}, and $1.7\%$ for \obj{2m0501} and $3\%$ for \obj{2m1425}.} Radial velocity (RV) and $v\sin(i)$ {values and their $1\sigma$ uncertainties are determined from their marginalized distributions obtained from our MCMC method.} For more detail on the method we refer to reader to \citet{Allers2016}. In Figure \ref{fig:2M0501_vsini} we show the observed spectrum of \obj{2m0501}, our fit to the data and the residuals of the fit. We show our measured $v\sin (i)$ and RV values for all three targets in Table \ref{tab:inclination}.

{\citet{Blake2010} have previously reported $v\sin(i)$ values for \obj{2m0045} and \obj{2m1425}. For \obj{2m0045}, they report a value of $32.82\pm0.17~$km/s, which is within $3\sigma$ of our value. Our obtained uncertainty ($\sim0.4$ km/s) is significantly larger than the reported uncertainty of $\sim0.17$ km/s in \citet{Blake2010}.
 For \obj{2m1425}, they report a $v\sin(i)$ value of $32.37\pm0.66~$km/s, which is within $2\sigma$ of our value. Our obtained uncertainties for \obj{2m1425} are similar to the reported uncertainties from \citet{Blake2010}. No previous $v\sin(i)$ values have been reported for \obj{2m0501}. }

By combining our measured $v\sin (i)$ values with a radius estimate \citep{Filippazzo2015}, we determine the maximum rotation period for each of our three targets. We find that the maximum rotation period for \obj{2m0045}, \obj{2m0501} and \obj{2m1425} are $\sim 6.4 $ hr, $19.6$ hr and $5.6$ hr respectively. In order to accurately constrain the rotation period of each of our targets we chose an observation duration twice that of each object's maximum rotation period for our \textit{Spitzer} variability monitoring observations.

\section{Spitzer Observations and Data Reduction} \label{sec:obs_datareduc}

We used the Infrared Array Camera \citep[IRAC;][]{Fazio2004} to observe our targets 
in the Channel 1 ($3.6~\mu$m) and Channel 2 ($4.5~\mu$m) bands as part of the Cycle 14 Program: ``Weather and Rotation of Young Brown Dwarfs'' (PID: 14019). 
We observed each target for twice their rotation period in both bands, resulting in a total observation duration of $12.8$ hr, $39.2$ hr and $11.2$ hr for \obj{2m0045}, \obj{2m0501} and \obj{2m1425} respectively.
The observations were designed following the recommendations for obtaining high precision photometry from the Spitzer Science Center. Science observations were preceded by a 30-minute dithered sequence to remove the initial slew settling that occurs when acquiring a new target, and followed by a 10-minute dithered sequence. The target was placed on the well-characterized ``sweet spot'' of the detector for science exposures to minimize correlated noise.

\begin{figure}[tb]
   \centering
   \includegraphics[scale=0.74]{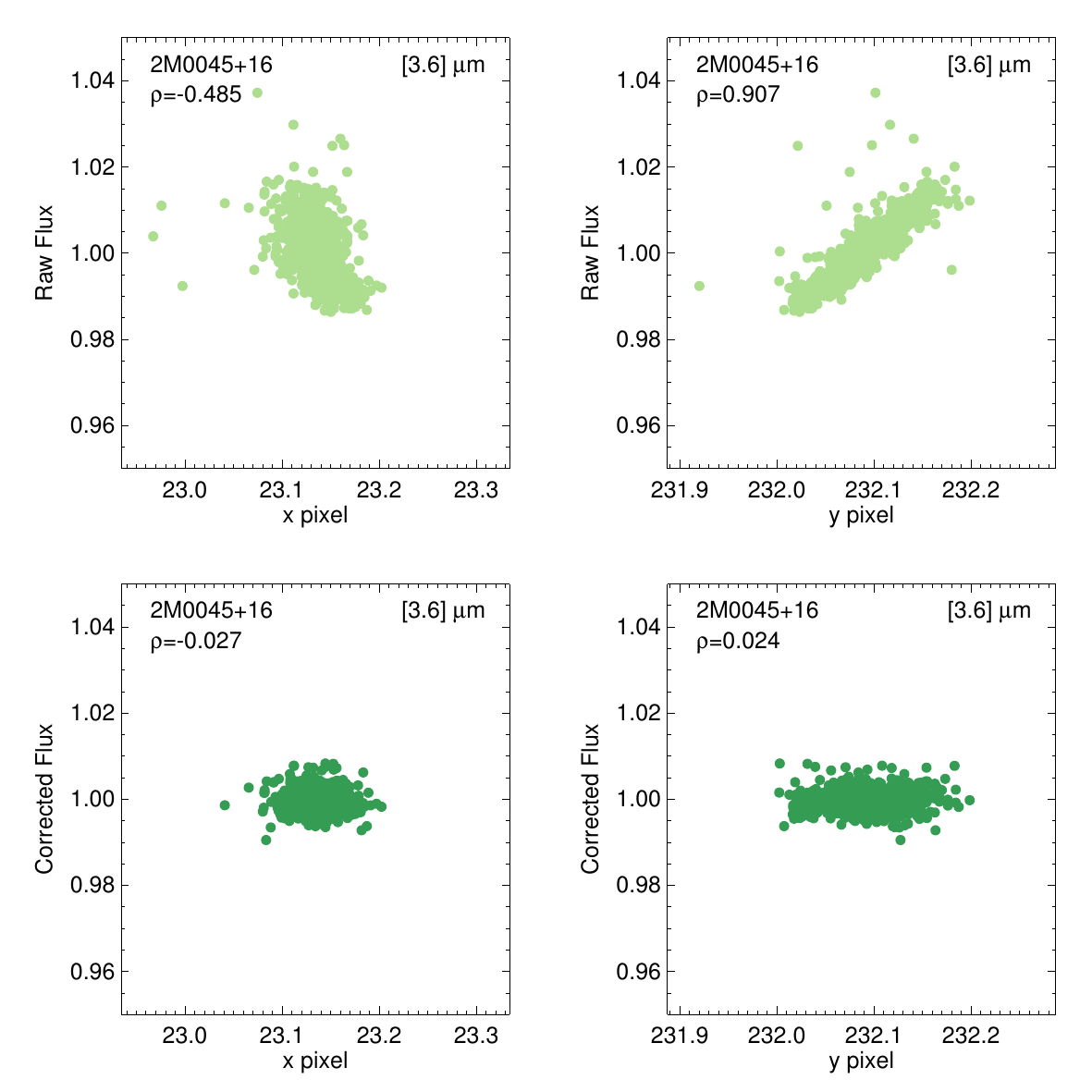}
     \includegraphics[scale=0.74]{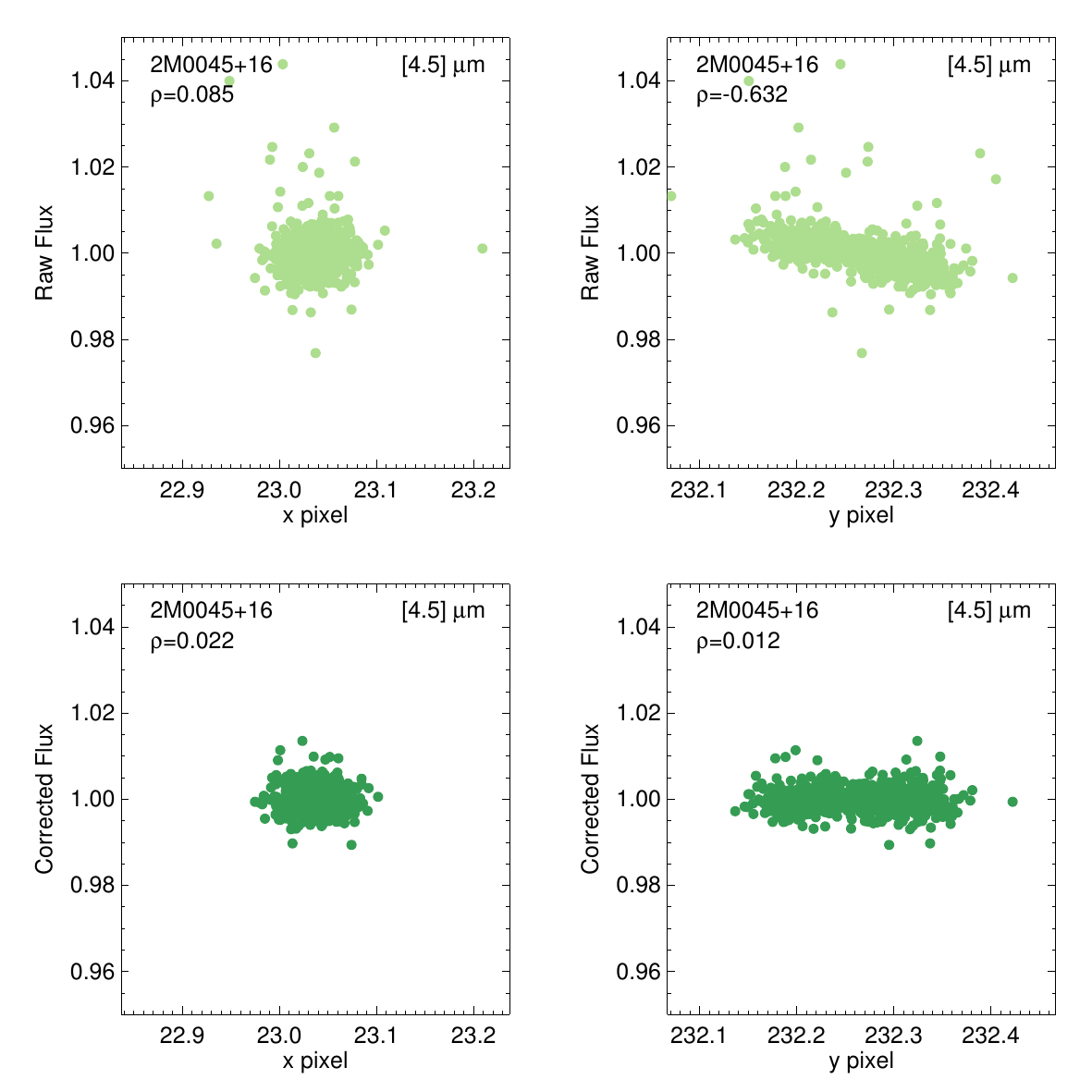}
   \caption{Raw \textit{Spitzer}/IRAC photometry obtained from aperture photometry is highly correlated with $x$ and $y$ sub-pixel positions. We correct for this by fitting a cubic function. Top panels show the correlation between the raw flux and pixel position. Bottom panels shows the corrected flux plotted against pixel position. Spearman's $\rho$ coefficient is a measure of the significance of correlation between two values -- in all cases Spearman's $\rho$ coefficient decreases after pixel phase correction. }
   \label{fig:2M0045_pixelphase}
\end{figure}

Photometry was obtained from the Basic Calibrated Data (BCD) images produced by the Spitzer Science Center using pipeline version S19.2. The {\sc{box\_centroider.pro}} routine was used to find the centroids of the target and reference stars of similar brightness in the field of view. Aperture photometry was performed on the target and reference stars using apertures with radii of $2.0-4.0$ pixels, in steps of $0.2$. We choose the final aperture size that produces the lowest rms target light curve. Outliers were identified and rejected from the raw light curves using a $6\sigma$ clip.

\begin{figure*}[tb]
   \centering
   \includegraphics[scale=0.75]{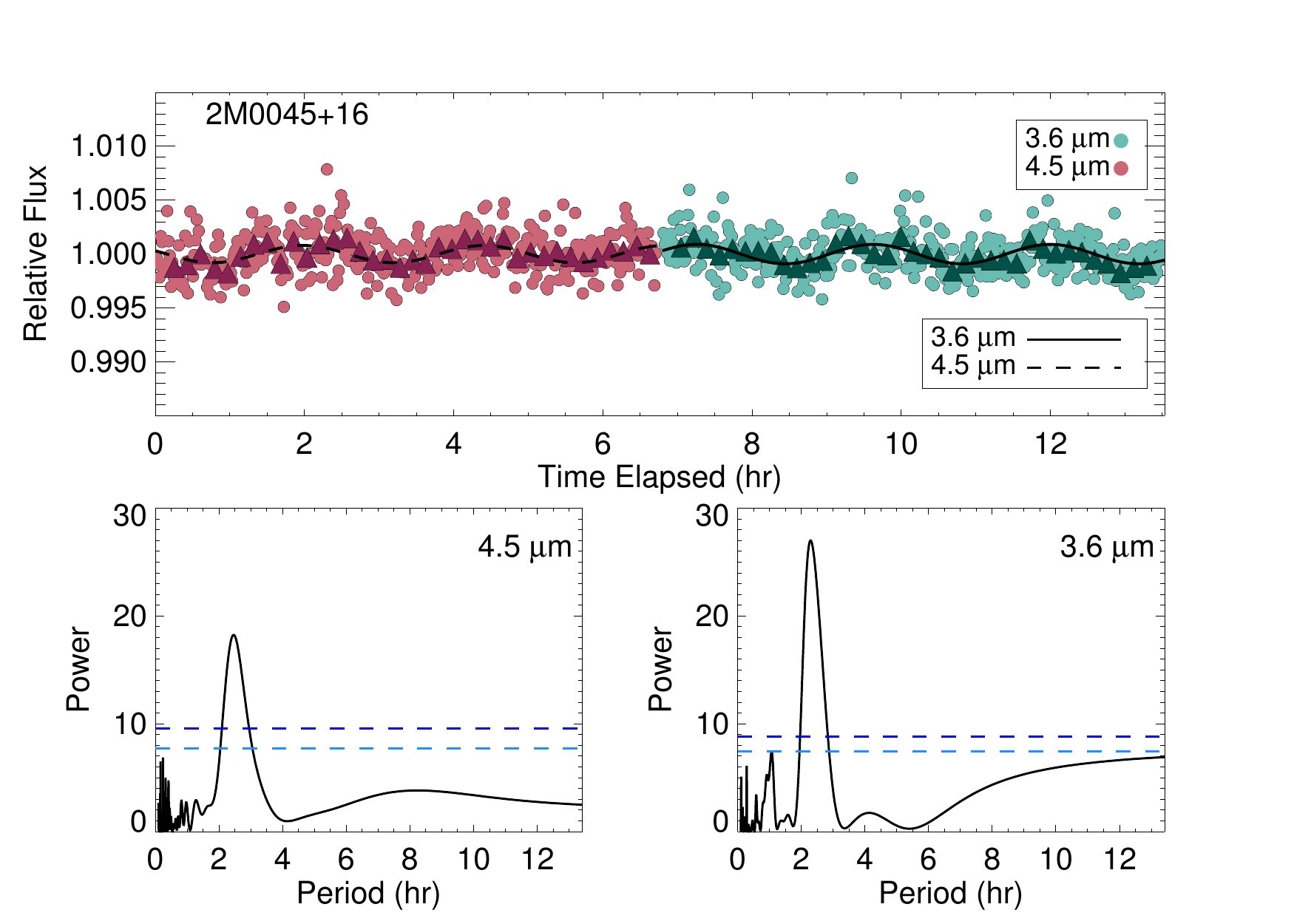}
   \caption{Top panel: Full corrected \textit{Spitzer} light curve of \obj{2m0045}. $3.6~\mu$m data is shown in green and $4.5~\mu$m is shown in pink. {{Circles} show 30-sec cadence and {triangles} symbols show 10 minute cadence.} Best-fit sinusoidal models from our MCMC analysis for each channel is overplotted in black. Bottom panels: Periodograms for each observation. The target periodogram is shown in black, reference star periodograms are shown in grey, and the $95\%$ and $99\%$ significance thresholds are shown by the blue dashed lines. The periodogram of \obj{2m0045} peaks well above the significance thresholds in both wavelengths. A sinusoidal model fit to both light curves favors a period of $2.4\pm0.1$ hr. }
   \label{fig:2M0045_full_lc}
\end{figure*}

\begin{figure*}[tb]
   \centering
   \includegraphics[scale=0.75]{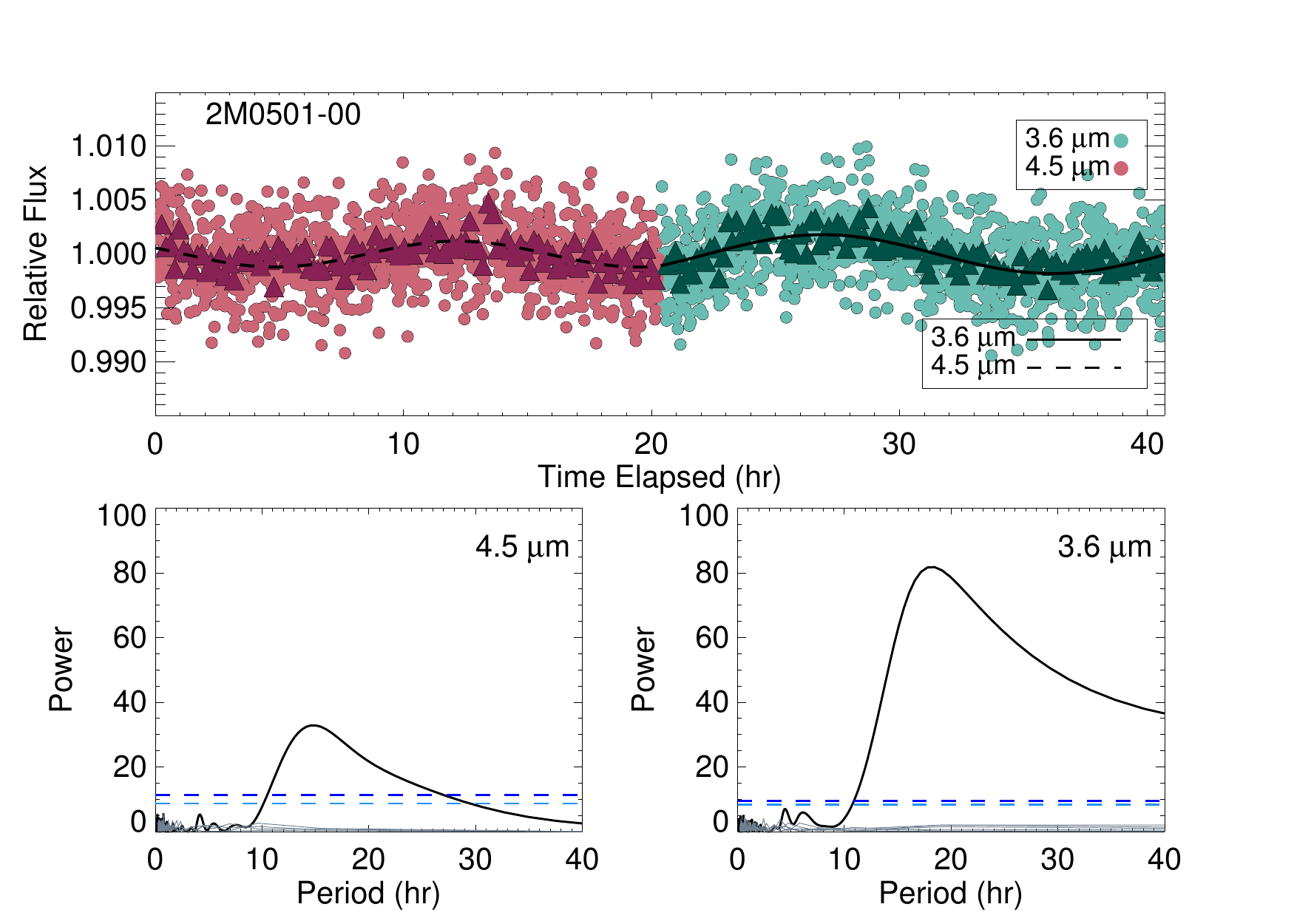}
   \caption{Top panel: Full corrected \textit{Spitzer} light curve of \obj{2m0501}. $3.6~\mu$m data is shown in green and $4.5~\mu$m is shown in pink. {{Circles} show 30-sec cadence and {triangles} symbols show 10 minute cadence.} Best-fit sinusoidal models from our MCMC analysis for each channel is overplotted in black. Bottom panels: Periodograms for each observation. The target periodogram is shown in black, reference star periodograms are shown in grey, and the $95\%$ and $99\%$ significance thresholds are shown by the blue dashed lines. The periodogram of \obj{2m0501} peaks well above the significance thresholds in both wavelengths. We fit a sinusoidal model to the full light curve, which favors a rotation period of $15.7\pm0.2$ hr.}
   \label{fig:2M0501_full_lc}
\end{figure*}

\begin{figure*}[tb]
   \centering
   \includegraphics[scale=0.75]{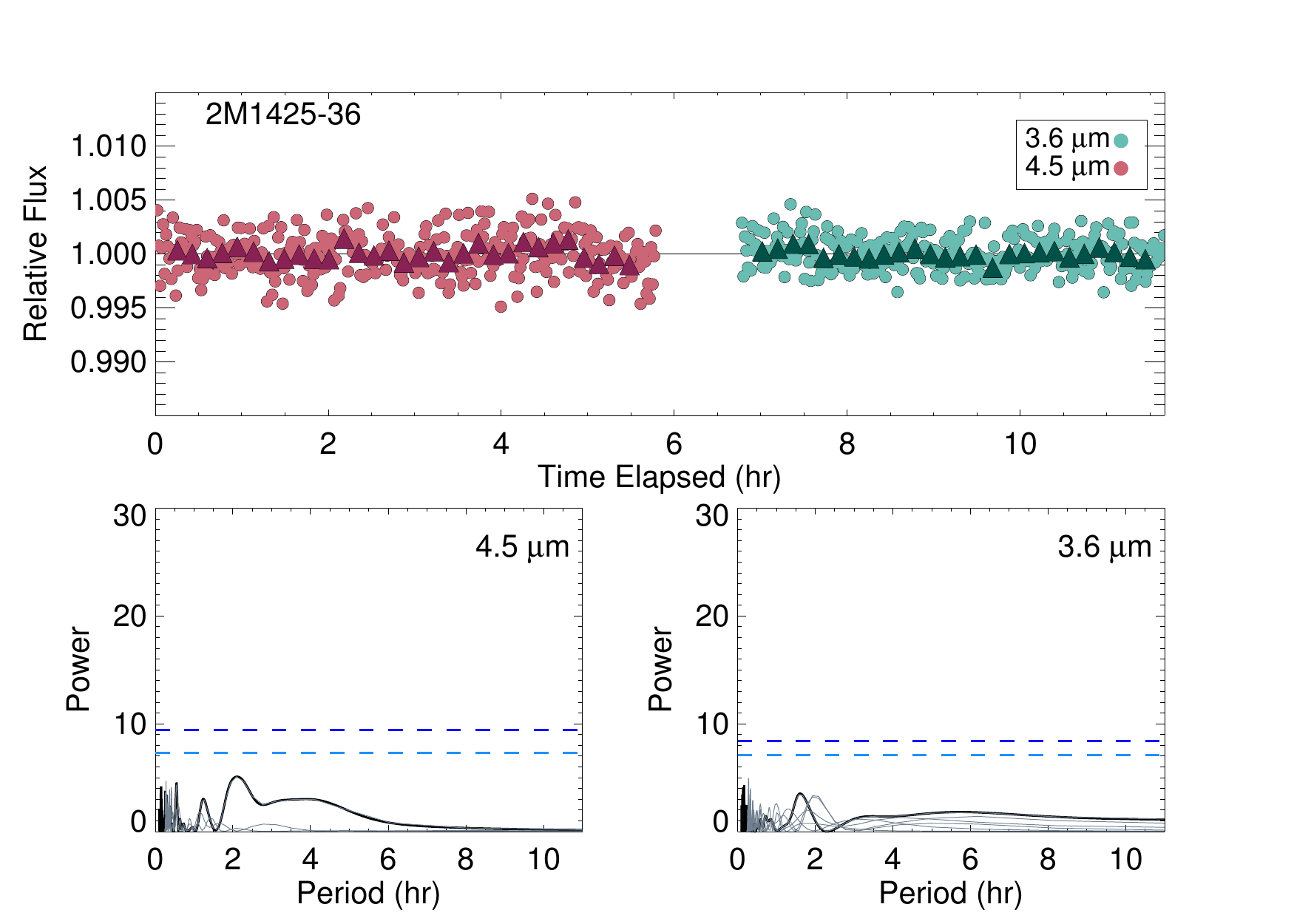}
   \caption{Top panel: Full corrected \textit{Spitzer} light curve of \obj{2m1425}. $3.6~\mu$m data is shown in green and $4.5~\mu$m is shown in pink. {{Circles} show 30-sec cadence and {triangles} symbols show 10 minute cadence.} Bottom panels: Periodograms for each observation. The target periodogram is shown in black, reference star periodograms are shown in grey, and the $95\%$ and $99\%$ significance thresholds are shown by the blue dashed lines. The periodogram of \obj{2m1425} does not show a significant peak in either observation.}
   \label{fig:2M1425_full_lc}
\end{figure*}

\textit{Spitzer}/IRAC photometry is known to exhibit a systematic effect due to intra-pixel sensitivity variations, known as the pixel phase effect. The top panels Figure \ref{fig:2M0045_pixelphase} show that the raw photometery is highly correlated with the $x$ and $y$ sub-pixel coordinates. We quantify the strength of correlation by calculating Spearman's $\rho$ coefficient \citep{Press1987}. We model the pixel phase effect as a function of the $x$ and $y$ coordinates, and find that linear and quadratic fits do not correct the observed correlation.
We model the pixel phase effect using a cubic function of the $x$ and $y$ coordinates \citep{Knutson2008, Heinze2015}:
\begin{equation}\label{eq:pixelphase}
    f(x,y) = P_0 + P_1x + P_2y + P_3 xy + P_4 x^2 + P_5 y^2 + P_6x^3 + P_7 y^3 + P_8x^2y + P_9 xy^2
\end{equation}
where $f(x,y)$ represents the measured flux, $P_i$ are the fitted coefficients, and $x$ and $y$ are the sub-pixel coordinates. 
{We correct the light curves of the target and reference stars using Equation \ref{eq:pixelphase}, and find that the fitted coefficients $P_i$ are similar for the target and reference stars.} We find that this correction decreases the correlation between the flux and pixel position for each observation (Figure \ref{fig:2M0045_pixelphase} for \obj{2m0045}). 
We present the final corrected light curves in the top panels of Figure \ref{fig:2M0045_full_lc}, \ref{fig:2M0501_full_lc} and \ref{fig:2M1425_full_lc}. {We additionally show how the $x$ and $y$ pixel positions vary during each observation in Figures \ref{fig:xypos_2M0045}, \ref{fig:xypos_2M0501} and \ref{fig:xypos_2M1425}. The variation in pixel position does not correlate with the corrected flux of our variable and non-variable targets.}

\section{Identification of Variables}\label{sec:var_id}
We search for variability in our observed light curves using the periodogram analysis outlined in \citet{Vos2018} and using a Bayesian framework outlined in \citet{Naud2017}.

\subsection{Variability Detection with Periodogram Analysis}
We calculate the Lomb-Scargle periodograms of our target and reference star light curves \citep{Scargle1982} to assess the significance of their trends. For each observation we calculate the $95\%$ and $99\%$ significance thresholds by simulating 1000 light curves from our observed reference stars. We create the simulated light curves by randomly rearranging the indices of the reference star light curves, which produces simulated light curves with Gaussian noise equal to that of our observed light curves. 
The $95\%$ and $99\%$ significance thresholds are plotted in blue in the bottom panels of Figures \ref{fig:2M0045_full_lc} - \ref{fig:2M1425_full_lc}. The periodograms of targets \obj{2m0045}, \obj{2m0501} (Figures \ref{fig:2M0045_full_lc} and \ref{fig:2M0501_full_lc}) display power that is significantly above the threshold at both $3.6~\mu$m and $4.5~\mu$m. The periodograms for each target peak at roughly the same period in both channels, which further supports that the variation is rotationally modulated. Our final target \obj{2m1425}, shown in Figure \ref{fig:2M1425_full_lc} does not exhibit periodogram power at the thresholds.



\begin{figure*}[tb]
   \centering
   \includegraphics[scale=0.55]{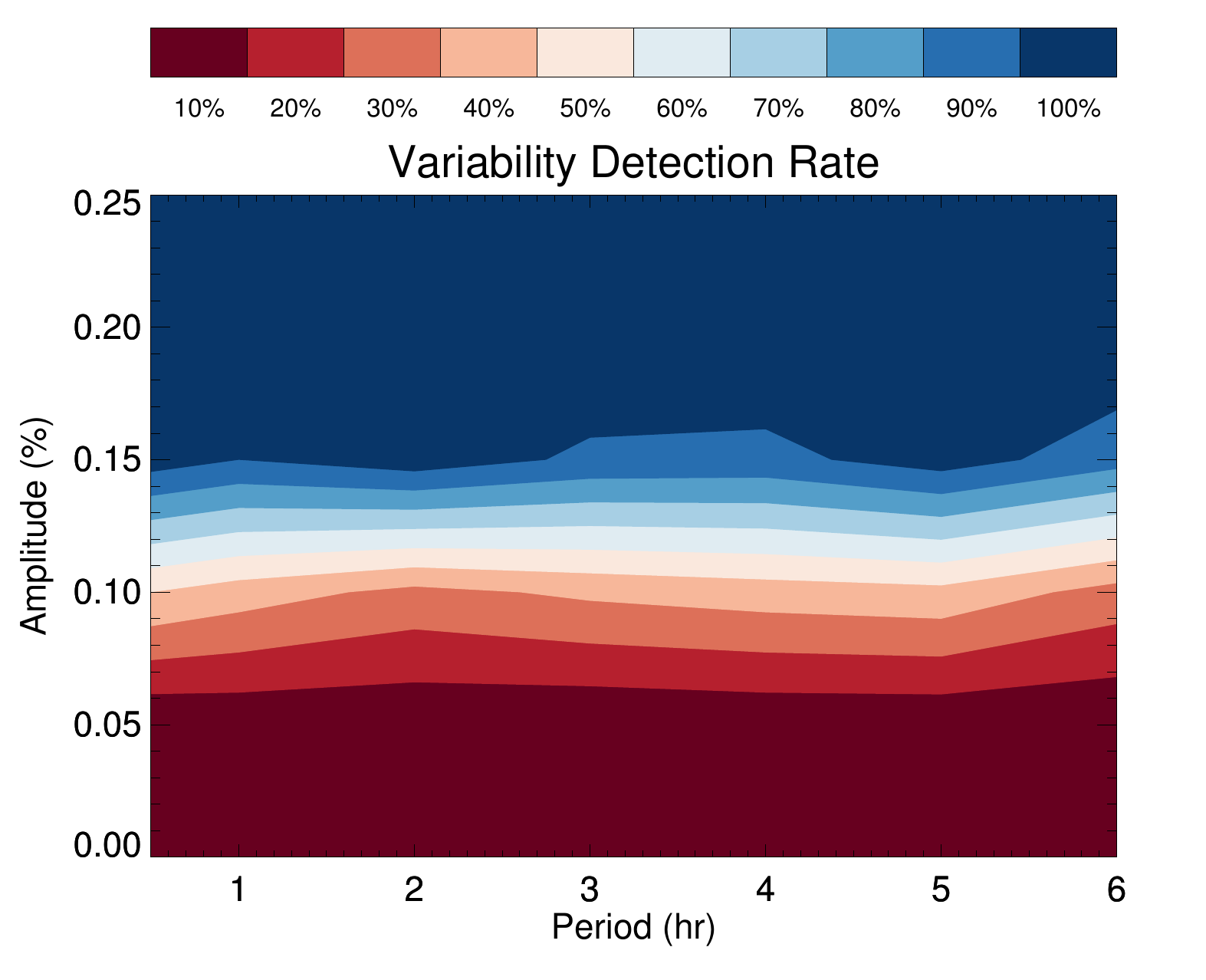}
    \includegraphics[scale=0.55]{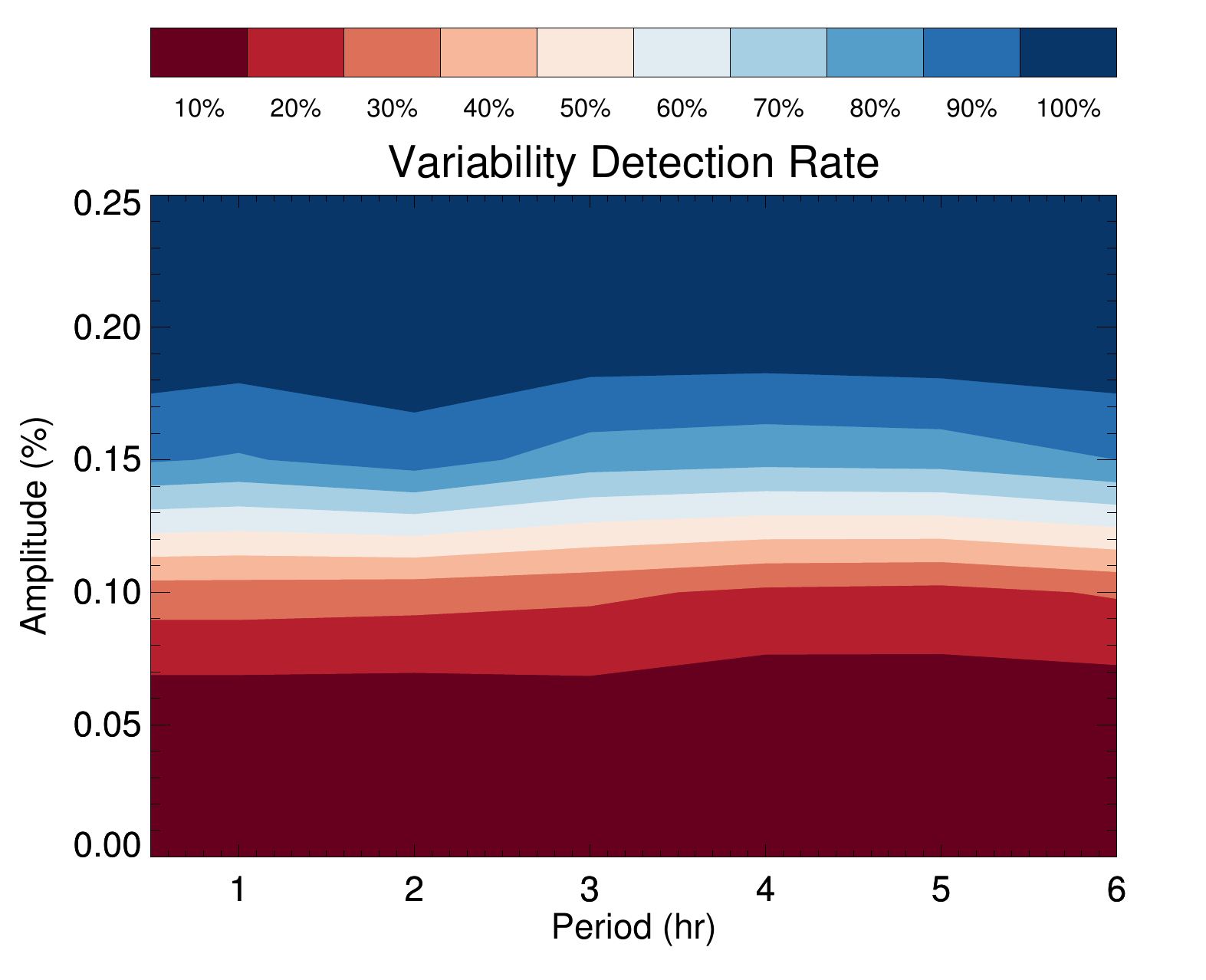}
   \caption{Sensitivity plots for \textit{Spitzer} $3.6~\mu$m (left) and $4.5~\mu$m (right) observations of \obj{2m1425}. The color scale shows the variability detection probability as a function of amplitude and period. Adopting a detection probability of $90
\%$ as our threshold, we place upper limits of $0.16\%$ and $0.18\%$ on the $3.6~\mu$m and $4.5~\mu$m variability amplitudes respectively.}
   \label{fig:sensitivity}
\end{figure*}

The periodogram analysis in Figure 
\ref{fig:2M1425_full_lc} confirms that \obj{2m1425} does not exhibit significant variability in either channel. The sensitivity of our \textit{Spitzer} observations and our knowledge of the maximum period for this target (5.6 hr, Section \ref{sec:rotvel}) allow us to place strong constraints on the upper limit of the variability amplitude at each wavelength. To determine these upper limits we create a sensitivity plot, which shows the variability amplitudes and rotation periods detectable by each observation. 
We inject sinusoidal curves into light curves with Gaussian-distributed noise similar to that of \obj{2m1425}. The simulated light curves have amplitudes of $0.05 - 0.5\%$, rotation periods of $0.5-6$ hr and random phase shifts. We analyse these simulated light curves using the periodogram analysis discussed above and calculate the detection probability as the percentage of light curves with a given variability amplitude and period that produces a periodogram power above the significance threshold. We show these sensitivity plots in Figure \ref{fig:sensitivity}. Adopting a detection probability of $90
\%$ as our threshold, we place upper limits of $0.16\%$ and $0.18\%$ on the $3.6~\mu$m and $4.5~\mu$m variability amplitudes respectively.

{Estimating significance of variability using periodogram analysis assumes that the noise properties of the target and reference stars are the same. This may not be the case if the reference stars are significantly brighter or fainter than the target. The observations of \obj{2m0501} include a number of reference stars with similar brightness, however \obj{2m0045} and \obj{2m1425} are brighter than their reference stars by $\Delta \mathrm{mag}\sim1-2$. The method also assumes that white noise is the only noise contribution. Reference stars with obvious variability are identified by eye and removed, but there may be residual time-correlated noise in the target and reference stars if the systematics were not adequately removed by Equation \ref{eq:pixelphase}. If this is the case the significance thresholds may be slightly underestimated. 
Since  none of the reference star periodograms in Figures \ref{fig:2M0045_full_lc}, \ref{fig:2M0501_full_lc} and \ref{fig:2M1425_full_lc} peak above the estimated significance thresholds and none of the reference star periodograms peak at periods similar to the periods of our variable targets, it is likely that the contributions from non-white noise do not affect our ability to identify variability using the Lomb-Scargle periodogram. Additionally, since periodic variability is independently recovered in both channels for \obj{2m0045} and \obj{2m0501}, it is likely that the variability is astrophysical in nature. }

{\subsection{Variability Detection Using the Bayesian Information Criterion} \label{sec:BIC}}

{We additionally use the Bayesian Information Criterion (BIC) to search for variability in each observation, following the method described in \citet{Naud2017}.  The BIC is defined as}
\begin{equation}
    \mathrm{BIC}=-2~\mathrm{ln}~\mathcal{L}_\mathrm{max} + k ~\mathrm{ln}~ N
\end{equation}
{where $\mathcal{L}_\mathrm{max}$ is the maximum likelihood achievable by the model, $k$ is the number of parameters in the model and $N$ is the number of datapoints used in the fit \citep{Schwarz1978}. }

{For each observation, we calculate $\Delta\mathrm{BIC} = \mathrm{BIC}_{sin} - \mathrm{BIC}_{\mathrm{flat}}$ to assess whether the variable sinusoidal or non-variable flat model is favored by the data. The BIC penalizes the sinusoidal model for having additional parameters compared with the flat model. These values are shown in Table \ref{tab:parameters}. A negative value of $\Delta\mathrm{BIC}$ indicates that the sinusoidal model is favored and a positive value indicates that the non-variable, flat model is favored. A $|\Delta\mathrm{BIC}|$ value between 0 and 6 indicates that one model is positively favored over the other, a value between 6 and 10 indicates that one model is strongly favored over the other and values above 10 indicate that one model is very strongly favored over the other \citep{Schwarz1978}. All of the $|\Delta\mathrm{BIC}|$ values shown in Table \ref{tab:parameters} are much greater than 10, showing that these results are highly significant. A variable, sinusoidal model is very strongly preferred for \obj{2m0045} and \obj{2m0501}, while a non-variable, flat model is very strongly favored for \obj{2m1425}. These results are fully consistent with the periodogram method for identifying variability discussed in Section \ref{sec:var_id}.}

\subsection{Determining rotation periods using MCMC}
To determine the rotation period and variability amplitude of our variable objects
we use the MCMC algorithm {\sc {emcee}} \citep{fm2013} to fit a sinusoidal model to the data in each band for \obj{2m0045} and \obj{2m0501}. Both variable objects exhibit fairly uniform, sinusoidal light curves and thus do not warrant a Fourier model fit with additional parameters \citep[e.g][]{Vos2018}. 
For the MCMC analysis we use 1000 walkers with 10000 steps. We discard the initial 1000 steps as the burn-in sample. We check for convergence by visually inspecting the resulting chains for each parameter to check that they are consistent with random noise of constant mean and variance. We also check that there is no difference between the parameter constraints obtained from the first and second halves of the chain. Based on these two checks, we find that the MCMC converges well for each sinusoidal model fit.

\begin{table*}[]
\begin{tabular}{lllllll}
\hline \hline
Target          &SpT    & Channel & Amplitude ($\%$)    & Period (hr)                 & Phase (rad)     & $\Delta$BIC     \\  [0.5ex]  \hline 
\obj{2m0045}~    &L2$\gamma$      & 1       & $0.18\pm0.04$       & $2.37^{+0.08}_{-0.06}$ & $1.2^{+0.9}_{-0.7}$ & $-44$                \\
                &        & 2       & $0.16\pm0.04$       & $2.43^{+0.09}_{-0.10}$ & $2.7\pm0.3$         &  $-28$            \\ \hline
\obj{2m0501}~    &L3$\gamma$      & 1       & $0.36\pm0.04$       & $18.5^{+0.8}_{-0.7}$   & $5.0\pm0.4$        &  $-212$               \\
                &        & 2       & $0.24\pm0.04$       & $14.7^{+0.9}_{-0.8}$   & $2.6\pm0.3$         & $-69$        \\    
                &        & 1\&2    & $0.28\pm0.02$       & $15.7\pm0.2$   & $3.1\pm0.1$                 &           \\    \hline  
\obj{2m1425}~    &L4$\gamma$      & 1       & $<0.16$ & $<5.6$    &   ...                                       &  $17$             \\
                &        & 2       & $<0.18$ &  $<5.6$   &    ...                                       &  $17$     \\  [0.5ex]  \hline  
\end{tabular}
\tablecomments {$\Delta$BIC $=\mathrm{BIC}_{\mathrm{sin}} - \mathrm{BIC}_{\mathrm{flat}} $}
\caption{Variability parameters of \obj{2m0045}, \obj{2m0501} and \obj{2m1425}. Values for \obj{2m0045} and \obj{2m0501} were obtaining using a sinusoidal model MCMC fit. Values for \obj{2m1425} were found from the sensitivity plots shown in Figure \ref{fig:sensitivity} and from the rotational velocity measurements discussed in Section \ref{sec:rotvel}. The $\Delta$BIC values show that the variable, sinusoidal model is strongly favored for \obj{2m0045} and \obj{2m0501} and the non-variable, flat model is strongly favored for \obj{2m1425}. }

\label{tab:parameters}
\end{table*}

We show the final posterior distributions of the variability amplitude, rotation period and phase in Figures \ref{fig:posterior_2M0045} and \ref{fig:posterior_2M0501}, and present the best-fit parameters and their 1-$\sigma$ errors in Table \ref{tab:parameters}. We overplot the best-fit sinusoidal model for each channel in Figures \ref{fig:2M0045_full_lc} and \ref{fig:2M0501_full_lc}. The residuals of the fit are normally distributed. 
For \obj{2m0045}, the measured rotation periods of $2.37^{+0.08}_{-0.06}$ hr at $3.6~\mu$m and $2.43^{+0.09}_{-0.10}$ hr at $4.5~\mu$m are fully consistent.
In contrast, the two rotation periods obtained for \obj{2m0501} are quite different -- $18.5^{+0.8}_{-0.7}$ hr at $3.6~\mu$m and $14.7^{+0.9}_{-.8}$ hr at $4.5~\mu$m. While the longer period fits the $3.6~\mu$m well, it does not provide a good fit to the $4.5~\mu$m data. The shorter period does a better job of fitting both channels. To estimate the most accurate rotation period for \obj{2m0501} we fit a sinusoidal model to both channels simultaneously, finding a period of $15.7\pm0.2~$hr. {We show the posterior distribution of the fit in Figure \ref{fig:posterior_2M0501_full}. }We adopt $15.7\pm0.2~$hr as the most likely rotation period.

{\subsection{Fitting the intra-pixel phase effect and astrophysical variability simultaneously}}
{The pixel phase effect can in principle be covariant with
astrophysical variability \citep{Heinze2014}. For the variable objects \obj{2m0045} and \obj{2m0501}, it is possible that in correcting for the pixel phase effect using Equation \ref{eq:pixelphase}, the variability signal may have been distorted. For these targets, we also fit their raw flux using a model that includes the pixel phase effect (Equation \ref{eq:pixelphase}) and a sinusoidal model to represent the variability, following previous \textit{Spitzer} brown dwarf variability and exoplanet transit studies \citep[e.g.][]{Metchev2015a,Delrez2018}.}

{We find that simultaneously fitting both the pixel phase effect and astrophysical variability simultaneously does not significantly affect the variability parameters in Table \ref{tab:parameters}. However, the simultaneous fit yields a higher correlation coefficient between the x and y-pixel positions and the measured flux in all cases, suggesting that including the sinusoidal fit during this step worsens the intra-pixel phase effect correction. We use the BIC framework discussed in Section \ref{sec:BIC} to assess whether the extra sinusoidal parameters are warranted by the data, and find that the pixel phase effect model (i.e. Equation \ref{eq:pixelphase}) is very strongly favored for each observation. This is likely because the pixel phase effect has a much larger effect on the photometry than the low-amplitude astrophysical variability. Since the results of the simultaneous fit are more highly correlated with pixel position than the original fit, we use the light curves obtained using the the cubic correction model (Equation \ref{eq:pixelphase}) for the rest of the analysis.}

{\section{Assessing the evidence for phase shifts between channels}}

We investigate the possibility of phase shifts between the $3.6~\mu$m and $4.5~\mu$m light curves. Different wavelengths probe different pressure levels in brown dwarf atmospheres \citep{Buenzli2012}, and phase shifts can potentially provide valuable information on the vertical atmospheric structure. Phase shifts have been observed in a number of brown dwarfs over the $1-5~\mu$m wavelength range \citep{Buenzli2012, Yang2016, Biller2018}, but have not been reported between \textit{Spitzer} Channels 1 and 2 \citep{Metchev2015a, Yang2016}. Since the $3.6~\mu$m and $4.5~\mu$m bands probe similar atmospheric pressures \citep{Yang2016}, phase shifts are not generally expected at these wavelengths.

{The BIC framework described in Section \ref{sec:BIC} also provides a robust method to determine whether the data warrant the addition of a phase shift between the $3.6~\mu$m and $4.5~\mu$m light curves for \obj{2m0045} and \obj{2m0501}. The two models allow the variability amplitude to change between the $3.6~\mu$m and $4.5~\mu$m light curves, but keep the period constant over both observations. The phase shift model includes an additional phase shift parameter for the $3.6~\mu$m data. The BIC is particularly useful in this case since it will penalise the phase shift model for having an additional parameter. For \obj{2m0501}, we calculated $\Delta \mathrm{BIC} = \mathrm{BIC}_{\mathrm{sin}} - \mathrm{BIC}_{\mathrm{phase shift}} = -25$, i.e. the sinusoidal model without a phase shift is strongly favored. For \obj{2m0045}, we find $\Delta \mathrm{BIC} = \mathrm{BIC}_{\mathrm{sin}} - \mathrm{BIC}_{\mathrm{phase shift}} = 5$. In this case the model that includes a phase shift between the light curves is positively favored, but this is not a significant result. We conclude that neither \obj{2m0045} nor \obj{2m0501} show a significant phase shift between their $3.6~\mu$m and $4.5~\mu$m light curves. }\\



\section{Comparison with Near-IR Variability Detections}
We previously detected variability in all three targets in our ground-based, $J$-band survey for variability in low-gravity brown dwarfs \citep{Vos2019}. Ground-based photometric monitoring is an excellent method for detecting variable objects in the near-IR, but due to shorter observation windows and weather constraints, space-based monitoring with \textit{Spitzer} is more effective at measuring rotation periods, particularly for low-gravity brown dwarfs which are thought to have longer rotation periods \citep[e.g.][]{Vos2018}. In \citet{Vos2019} we detected significant $J$-band variability in \obj{2m0045} during two $\sim4$ hr epochs separated by 2 years. We measured an amplitude of $\sim1\%$ during both observations, and found no evidence for light curve evolution between the two epochs. Our \textit{Spitzer} light curve appears sinusoidal over the entire observation, so it seems that both the $J$-band and mid-IR light curves are stable.
Periodogram analysis of the ground-based $J$-band light curve suggested a rotation period of $3-6$ hr \citep{Vos2019}, while in this paper we measure a period of $2.4\pm0.1$ hr. Both of the ground-based, $J$-band light curves have significant gaps in the data due to poor weather during the observation, and this may explain the discrepancy between the estimated ground-based $J$-band and measured \textit{Spitzer} rotation period.

\begin{figure*}[tb]
   \centering
   \includegraphics[scale=0.78]{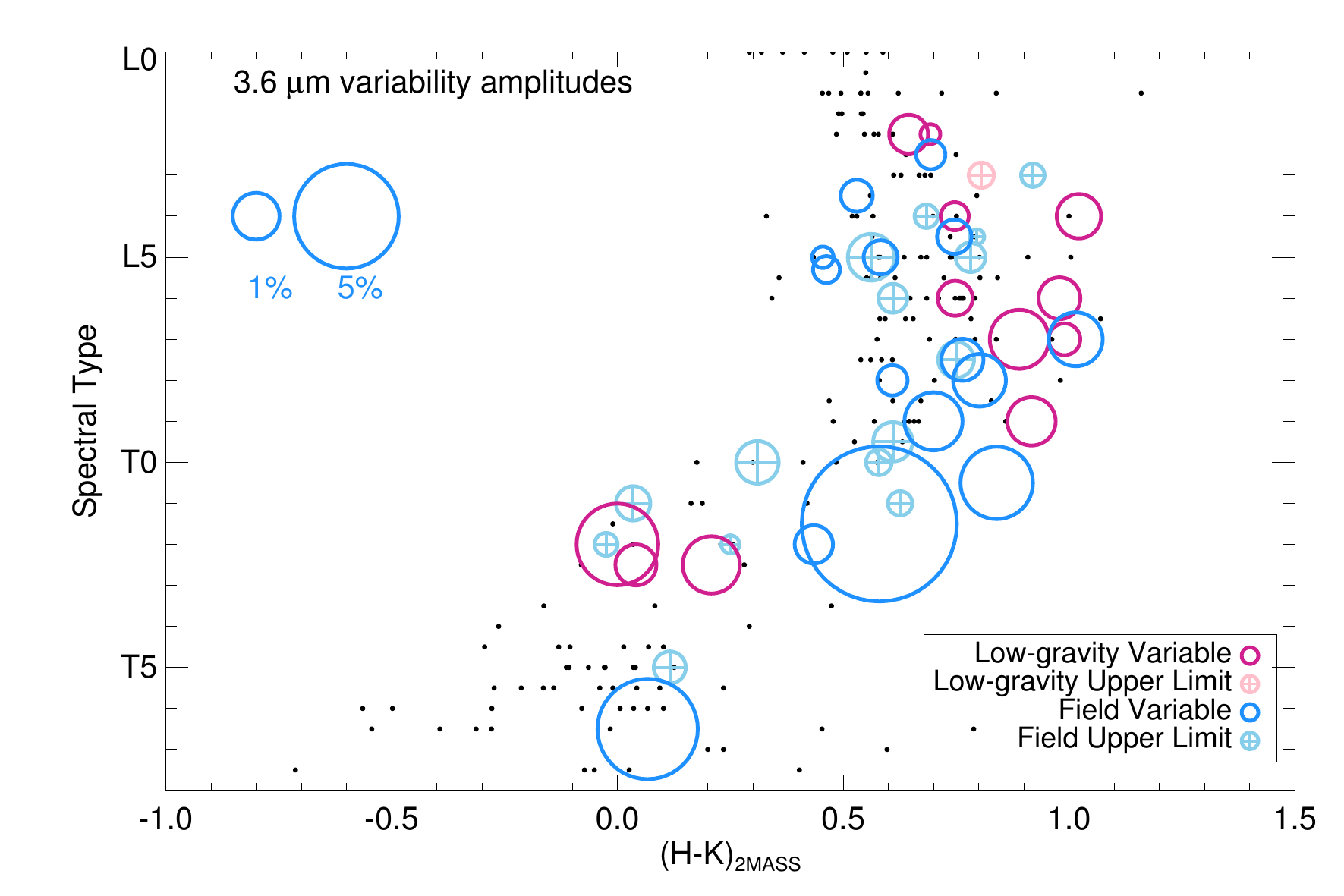}
    \includegraphics[scale=0.78]{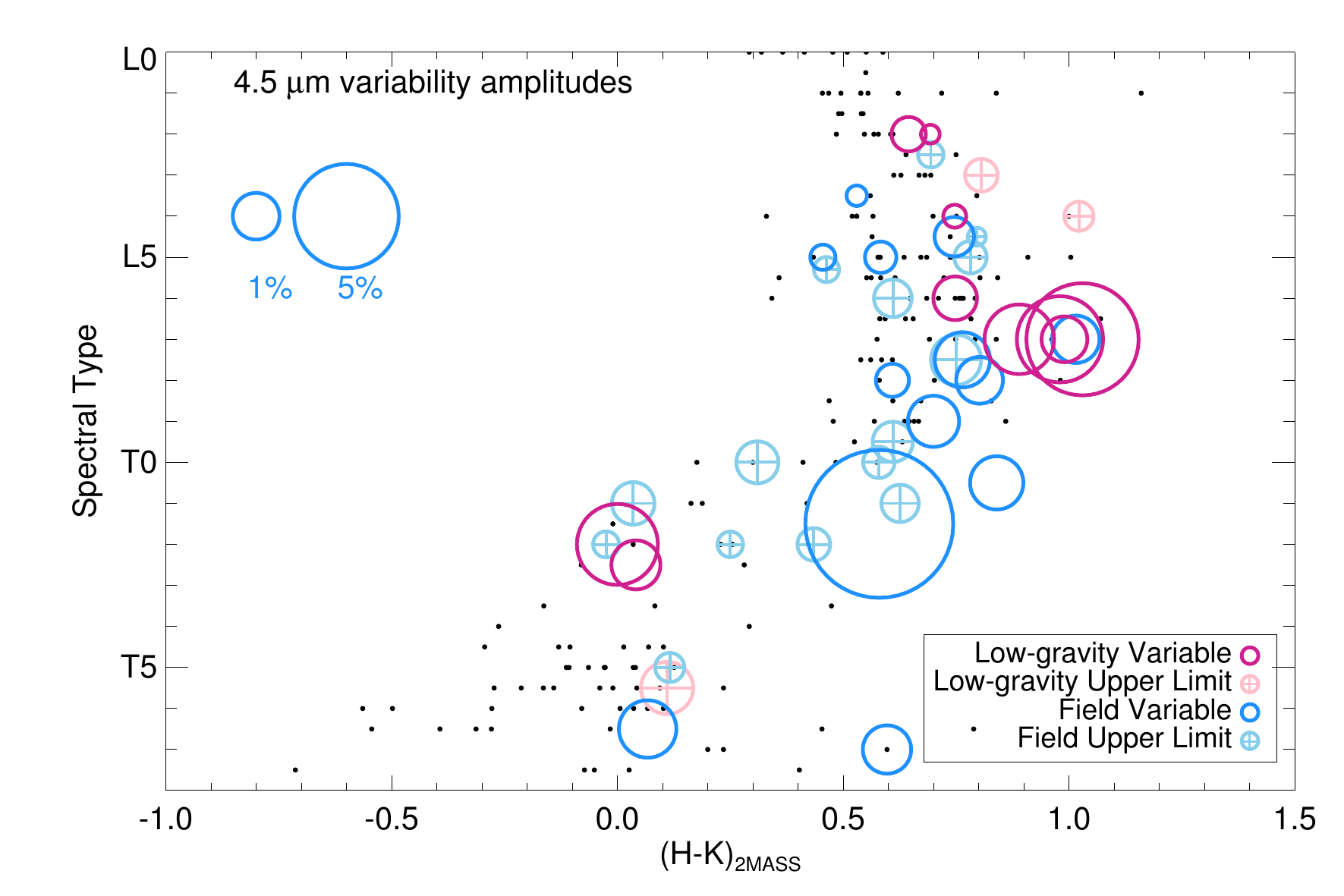}
   \caption{Brown dwarfs with \textit{Spitzer} detections of variability as a function of spectral type and $(H-K)_{2MASS}$ color. Black points show the field L-T dwarf sequence from \citet{Dupuy2012}. The sample of field brown dwarfs with detected mid-IR variability is shown by the dark blue circles and the sample of young mid-IR variables is shown in  dark pink. {Variability amplitude upper limits are shown for the field and young population in light blue and light pink circles with crosses respectively.} The symbol area is proportional to the variability amplitude. The data used for this figure is shown in Table \ref{tab:variables}.}
   \label{fig:amplitudes}
\end{figure*}

We detected $J$-band variability in \obj{2m0501} in two epochs in \citet{Vos2019}. During each $\sim4$ hr observation, we detected a slow downward trend, with a maximum $J$-band amplitude of $\sim2\%$. Since we did not cover a full rotation period in either observation, our periodogram analysis constrained the period to $>5$ hr, which is consistent with our much longer \textit{Spitzer} measurement of $15.07\pm0.2$ hr. \obj{2m0501} exhibits higher $J$-band and \textit{Spitzer} mid-IR amplitudes than \obj{2m0045}.

\obj{2m1425} was the lowest amplitude variable presented in \citet{Vos2019}. We detected variability during one epoch only, with an amplitude of $\sim0.7\%$. Our ground-based $J$-band monitoring did not cover a full rotation period and our periodogram analysis favoured periods of $2-4$ hr \citep{Vos2019}. We do not detect mid-IR variability in this object, and place upper limits on the variability amplitude in each channel. Comparing the near-IR and mid-IR amplitudes of all three L dwarfs observed in this paper reveals that \obj{2m0501} shows the highest mid-IR and near-IR amplitudes, followed by \obj{2m0045}, and \obj{2m1425} shows the lowest amplitudes. All three targets show smaller amplitudes in the mid-IR than the $J$-band , which is consistent with such amplitude measurements in young and field brown dwarfs \citep{Biller2018, Metchev2015a}. Atmospheric models predicts that the $J$-band probes deeper pressure levels than mid-IR wavelengths, which would explain the lower mid-IR amplitudes \citep{Buenzli2012}.

\section{Mid-IR Variability Amplitudes of Young L Dwarfs} \label{sec:amplitudes}

Observed variability amplitudes are thought to vary with spectral type \citep{Metchev2015a, Radigan2014}, inclination angle \citep{Vos2017} and surface gravity \citep{Metchev2015a, Vos2019}. In Table \ref{tab:variables} we show the full sample of brown dwarfs with measured variability amplitudes and/or upper limits on infrared variability. This table includes their infrared amplitudes, rotation periods, estimated ages and inclination angles. 
Figure \ref{fig:amplitudes} shows the full sample of field brown dwarfs and young brown dwarfs with measured \textit{Spitzer} variability amplitudes on a spectral type color diagram. The area of the symbol size is proportional to the observed variability amplitude. {Upper limits on the variability amplitudes of brown dwarfs are shown by lighter colored circles with a cross.  }With the addition of \obj{2m0045} and \obj{2m0501} at early L spectral types we can begin to study the variability properties of the low-gravity population across the entire L sequence. While the sample is still relatively small some tentative trends emerge.

\begin{figure*}[tb]
   \includegraphics[scale=0.72]{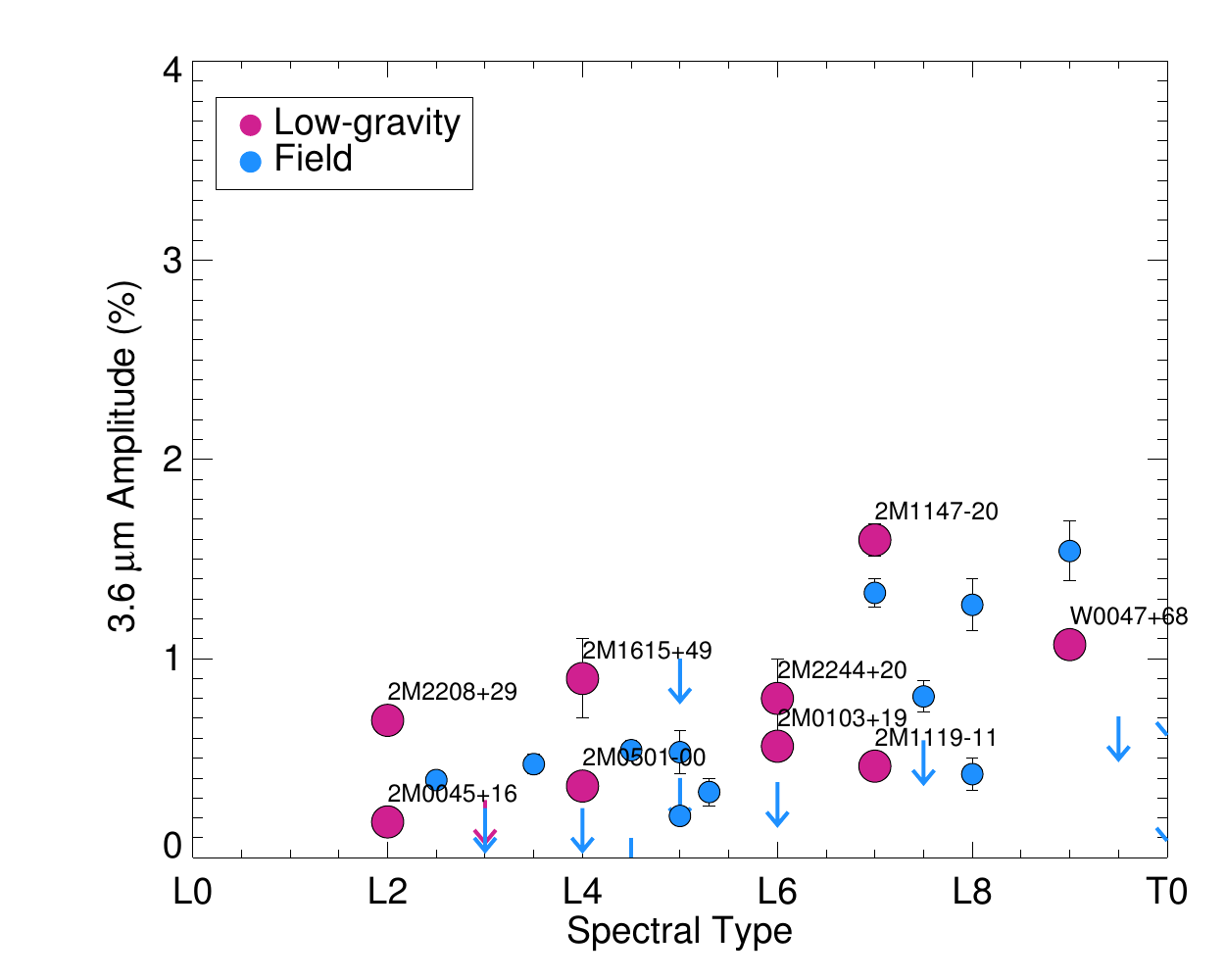}
   \includegraphics[scale=0.72]{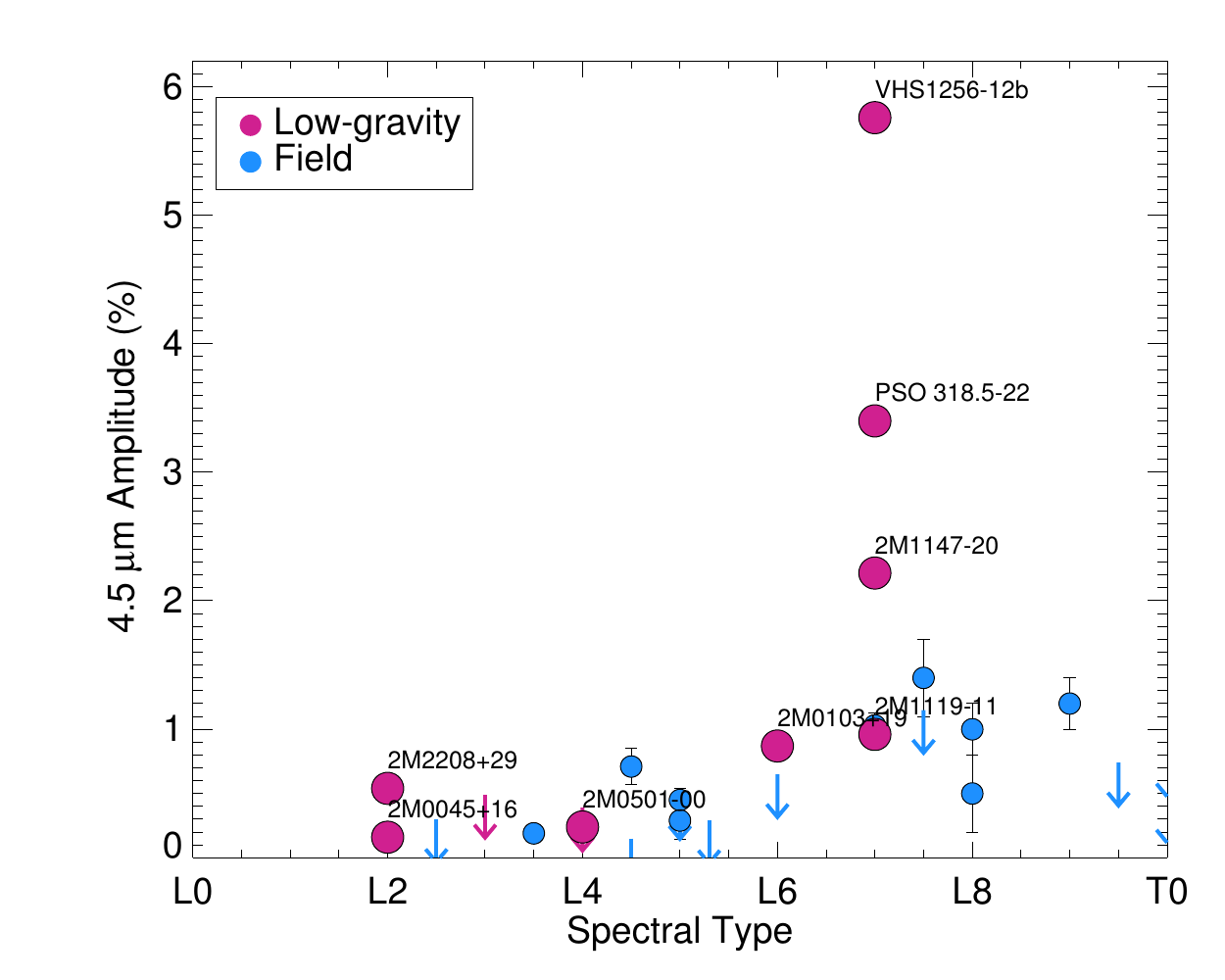}
   \caption{\textit{Spitzer} peak-to-peak variability amplitudes as a function of spectral type for L dwarfs. {Variability amplitude upper limits are shown for the field and young population by downward pointing arrows .} Low-gravity $3.6~\mu $m detections (left) follow the same trend as the field brown dwarfs -- increasing amplitude with cooler spectral type. However the $4.5~\mu $m detections show a tentative enhancement in amplitude at late-L spectral types for low-gravity objects only.}
   \label{fig:amplitudes_plot}
\end{figure*}

The left panel of Figure \ref{fig:amplitudes_plot} shows the measured $3.6~\mu$m variability amplitudes as a function of spectral type. It is apparent that the maximum amplitudes for both field and young dwarfs increase with cooler spectral type in the L sequence, as noted by \citet{Metchev2015a} for the field dwarf population. \citet{Metchev2015a} also find a tentative correlation between low-gravity and high-amplitude variability for the low-gravity objects with spectral types L3-L5.5. However, the current sample of low-gravity variables show similar $3.6~\mu$m amplitudes to the field dwarf population. It is worth noting that for each spectral type, the object with the highest measured amplitude is young -- so there is the possibility that the highest intrinsic amplitudes occur in young objects, and that the observed amplitudes can be reduced by secondary effects such as inclination angle \citep{Vos2017}.

   \hspace*{-40pt}

The right panel of Figure \ref{fig:amplitudes_plot} shows the amplitudes of $4.5~\mu$m variability detections as a function of spectral type. For early-L spectral types, the young population shares similar amplitudes with the field brown dwarfs. In the late-L population, the four young $\sim$ L7 objects PSO J318.5-22 \citep{Biller2018}, WISEA 1147-2040, 2MASS 1119-1137AB \citep{Schneider2018} and VHS1256-12b \citep{Bowler2020,Zhou2020} show enhanced amplitudes compared to the field population. Note that since 2MASS 1119-1137AB is a binary system \citep{Best2017}, its peak-to-peak amplitude of $0.96\%$ is likely underestimated since the variability signal may be diluted by its unresolved companion. 

In the near-IR there seems to be a distinction in the behaviour of late-L young brown dwarfs compared to field brown dwarfs. While older mid to late-L dwarfs \citep{Yang2016, Manjavacas2018} show a linear amplitude dependence on wavelength, the young objects PSO J318.5$-$22 \citep{Biller2018}, WISEP J004701.06$+$680352.1 \citep[W0047, ][]{Lew2016} and  HD203030B \citep{Miles-Paez2019} show different amplitudes in the water absorption band at $1.4~\mu$m. PSO J318.5$-$22 and W0047 both show decreased amplitudes in the water band while HD203030B shows a marginal enhancement in amplitude in the water band. Furthermore, the water band variability amplitude of PSO J318.5-22 appears to change between two rotations - initially showing a suppressed amplitude followed by an amplitude similar to the continuum amplitude the following rotation \citep{Biller2018}.  \citet{Miles-Paez2019} suggest that this may be due to an increased height differentiation between the condensate cloud layers and the high altitude water and carbon monoxide layer in low-gravity objects. This would result in higher variability amplitudes for the young objects in wavelength regions that are relatively free of water and carbon monoxide gas species. Enhanced variability amplitudes have indeed been observed in the $J$-band \citep{Vos2019}. In the mid-IR however, we tentatively observe enhanced amplitudes in late-L low-gravity objects at $4.5~\mu$m but not at $3.6~\mu$m. More extensive mid-IR monitoring programs will be essential to statistically compare the amplitudes between the two populations at $3.6~\mu$m and $4.5~\mu$m.

\begin{figure*}[tb]
   \centering
   \includegraphics[scale=0.8]{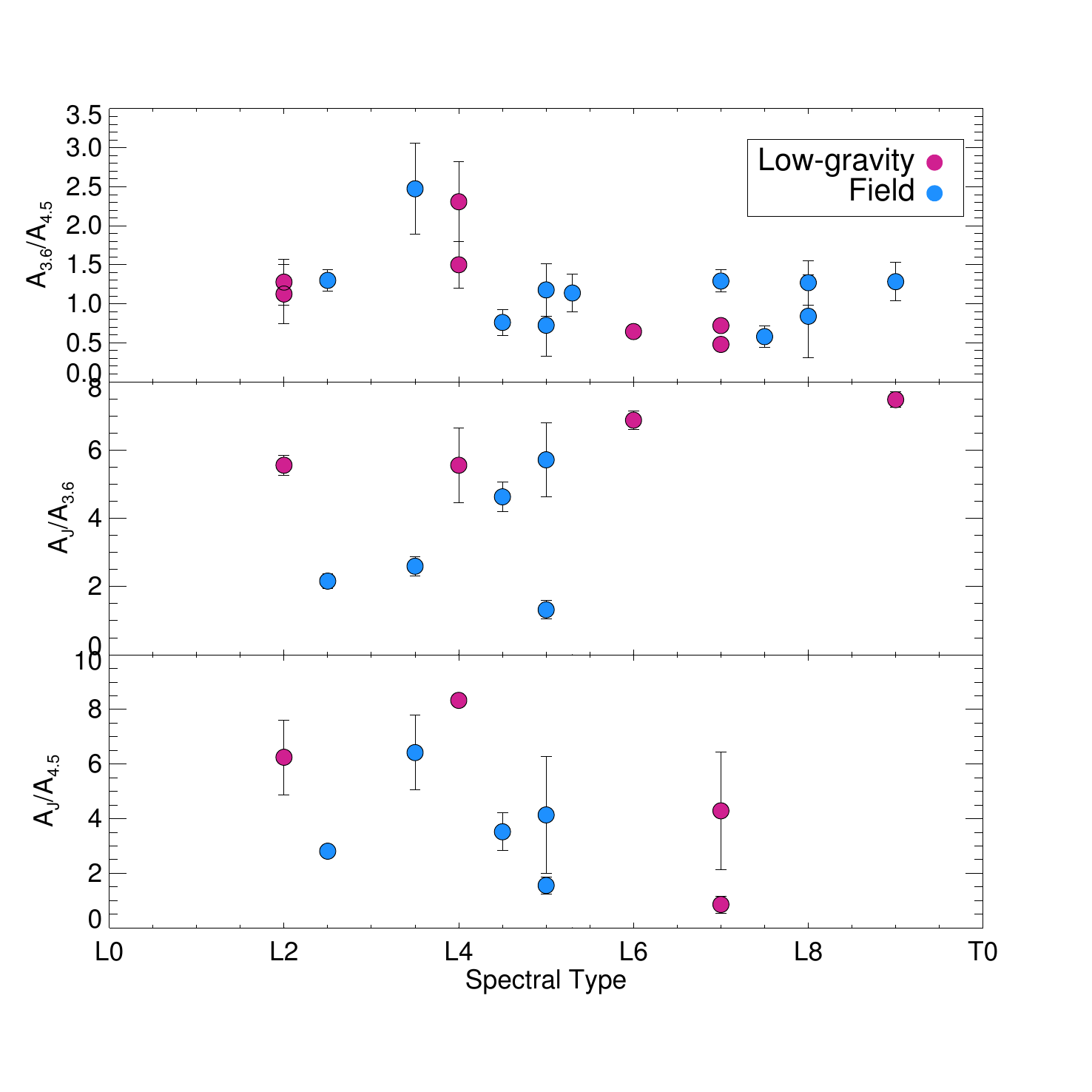}
   \caption{Variability amplitude ratios of field and young brown dwarfs. We do not find any significant correlations between variability amplitudes in the $J$-band, at $3.6~\mu$m and $4.5~\mu$m for young and field dwarfs.}
   \label{fig:amplitude_ratios}
\end{figure*}

Figure \ref{fig:amplitude_ratios} shows variability amplitude ratios ($A_{J}, A_{3.6}, A_{4.5}$) measured for low-gravity and field L dwarfs. The brightness temperature of L dwarf atmospheres is strongly dependent on wavelength, and the variability amplitude ratios in the near and mid-IR have been used to estimate the temperature gradient among cloud layers or between regions of thin and thick clouds \citep{Radigan2012, Heinze2014}. \citet{Metchev2015a} find no correlation between spectral type and $A_{3.6}/A_{4.5}$ amplitude ratio. 
The top panel of Figure \ref{fig:amplitude_ratios} shows that the new sample of young objects are not obvious outliers in this plot. The mean $A_{3.6}/A_{4.5}$ across the L0-T0 range is $1.2\pm0.5$ for field brown dwarfs and $1.2\pm0.6$ for low-gravity objects, thus we see no difference between the two populations.
The bottom two panels of Figure \ref{fig:amplitude_ratios} show ratios involving $J$-band detections. Apart from PSO J318.5$-$22 \citep{Biller2018}, the mid-IR and $J$-band measurements were not taken simultaneously. Since variability amplitudes are known to evolve rapidly in some cases \citep[e.g.][]{Apai2017}, these results should be interpreted with a degree of caution. 
The middle panel, which shows $A_{J}/A_{3.6}$ amplitude ratio, by eye suggests that this ratio may be higher for the young population.
The mean $A_{J}/A_{3.6}$ amplitude ratio is $3.3\pm1.8$ for the field dwarfs and $6.4\pm1.0$ for the low-gravity objects. While the mean amplitude ratios in this case are significantly different, more simultaneous amplitude measurements are necessary to robustly investigate this possible trend.
Finally, the bottom panel shows the $A_{J}/A_{4.5}$ amplitude ratios for variable L dwarfs. The mean amplitude ratio is $3.7\pm1.8$ for the field dwarfs and $4.9\pm3.2$ for the low-gravity sample, so we find no evidence of a difference between the samples. Future simultaneous variability observations with \textit{JWST}/NIRcam in the long and short wavelength channels may shed light on the amplitude ratios of variability brown dwarfs in the future.

\section{Inclination Angles of Variable Brown Dwarfs} \label{sec:inclination}

Combining our measurements for the rotation period and rotational velocity with a radius estimate allows us to place constraints on the inclination angles of our targets. 
\citet{Filippazzo2015} provide radius estimates for all three targets. They combine their calculated bolometric luminosity ($L_\mathrm{bol}$) with the inferred age of each target to find the range of predicted radii from evolutionary models. Radii estimates, which depend on gas and condensate chemistry, molecular opacities, cloud modelling an atmospheric boundary conditions \citep{Saumon2008}, can be heavily dependent on the models. To address this, \citet{Filippazzo2015}
use the solar metallicity, hybrid cloud (SMHC08) models \citet{Saumon2008}, the DUSTY00 models \citep{Chabrier2000} and the $f_\mathrm{sed}=2$ (SMf208) models \citep{Saumon2008} to estimate the radii. Their final radius range for each source is the minimum and maximum values of all three model predictions for the given age and $L_\mathrm{bol}$, and are not formal 1$\sigma$ uncertainties. The range of estimated radii for the field brown dwarfs are in good agreement with the handful of directly measured field dwarf radii \citep[e.g.][]{Pont2005, Deleuil2008, Bouchy2010, Siverd2012, Littlefair2014}, however there has not yet been an empirical test of the radii of the young ($<400~$Myr) objects. Thus, for the young objects in particular it is possible that the radii, and therefore the inclinations may be biased in some way.

We use Monte Carlo analysis to calculate the inclination of each target using normal distributions for the periods and radii, and the $v \sin (i)$ distributions obtained in Section \ref{sec:rotvel}. $\sin (i)$ values that fell above 1 were set equal to 1 since discarding them biased the results to lower inclination angles. The inclination and error were calculated from the median and standard deviation of the resulting distribution of $\sin (i)$. Since we did not detect significant variability for \obj{2m1425}, we do not have a measured rotation period. However, we can set a lower limit on the rotation period of 2.5 hr based on our ground-based variability detection \citep{Vos2019}, and an upper limit from our $v\sin (i)$ measurement (Section \ref{sec:rotvel}). We compute the inclination angle of \obj{2m1425} using the method described above, but we used a uniform distribution of $2.5-5.6$ hr for the rotation period based on our $J$-band light curve published in \citet{Vos2019}, and the maximum period determined from our $v\sin(i)$ measurement.

We show our input values for $v \sin (i)$, period and radius and our resulting inclination angles in Table \ref{tab:inclination}. We find that \obj{2m0045} has an inclination angle of $22\pm1^{\circ}$, placing it close to a pole-on alignment. We calculate an inclination angle of ${60^{+20 }_{-9}} ^{\circ}$ for \obj{2m0501}. This inclination is less accurate due to the large error bar on the rotation period, but \obj{2m0501} is clearly closer to equator-on than \obj{2m0045}. We find that \obj{2m1425} is inclined with an angle of ${54^{+36}_{-15}}^{\circ}$, however a rotation period measurement is necessary to confirm this inclination angle.

\begin{figure}[tb]
   \centering
   \includegraphics[scale=0.7]{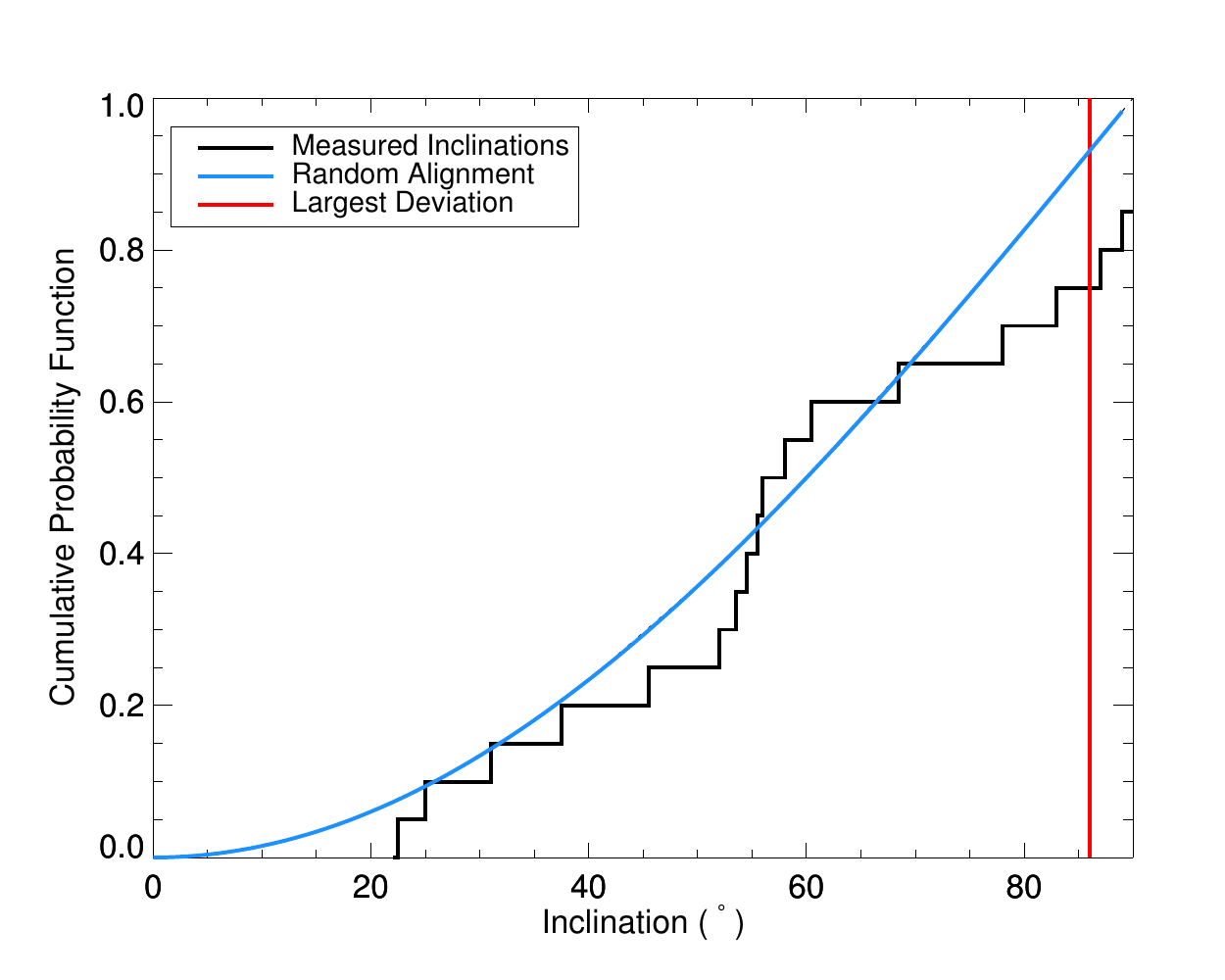}
   \caption{{The cumulative probability distribution for a sample of objects with random orientation (blue) compared to the measured sample of brown dwarf inclination angles (black). The Kolmogorov Smirnov test results in a maximum deviation of $\sim0.18$ at $86^{\circ}$, and a significance of 0.5. Our results show that there is no significant difference between the two distributions.}}
   \label{fig:CDF}
\end{figure}

\subsection{Are the inclination angles randomly aligned?}
\begin{figure}[tb]
   \centering
   \includegraphics[scale=0.7]{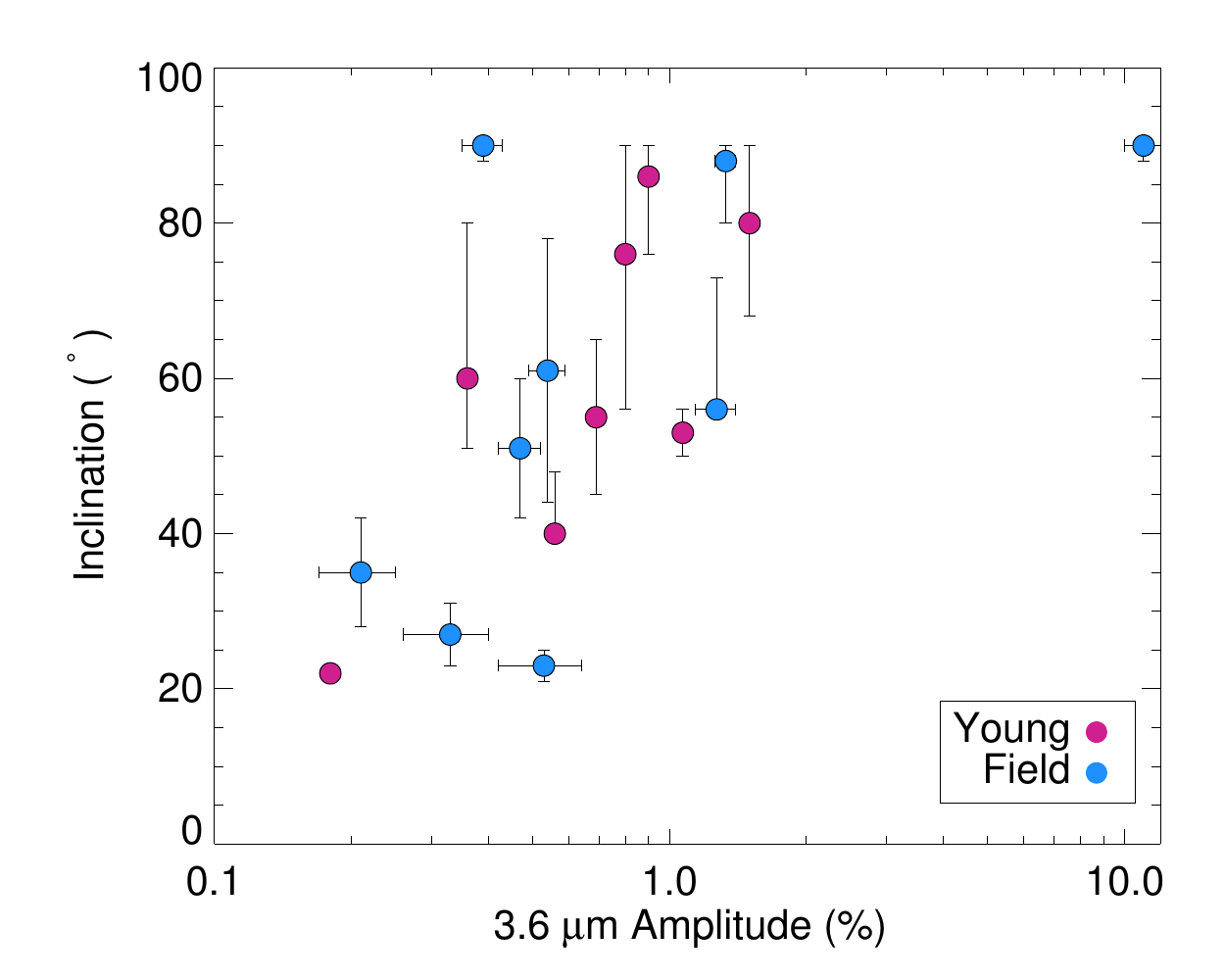} 
       \includegraphics[scale=0.7]{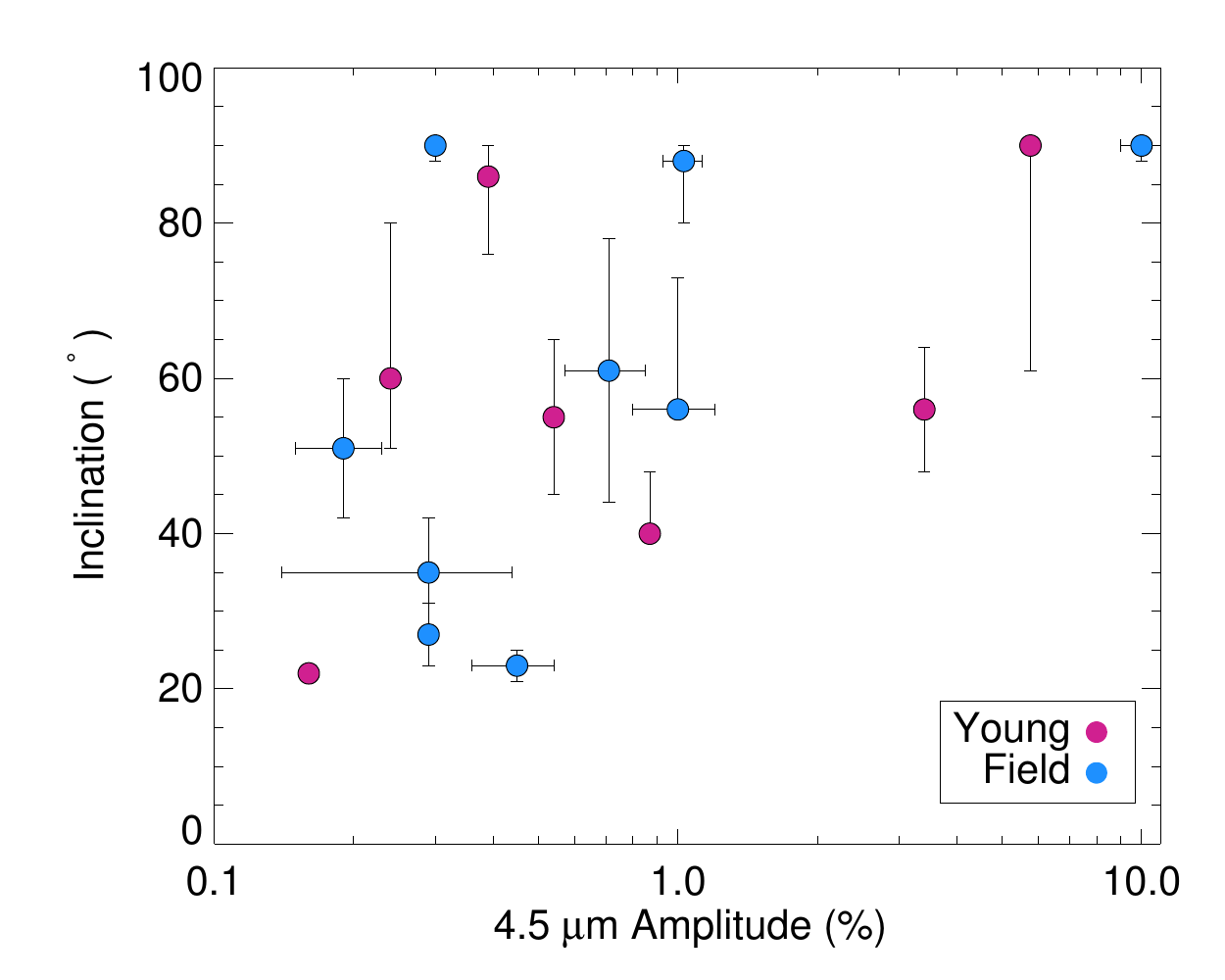}
   \caption{Inclination angle plotted against \textit{Spitzer} [3.6] (left) and [4.5] (right) variability amplitude. The population of high-amplitude variables are viewed close to equator-on ($90^{\circ}$), while the maximum variability amplitudes decrease as the object is viewed closer to pole-on ($0^{\circ}$).}
   \label{fig:inclination_amplitude}
\end{figure}

{Our total sample of 18 objects with measured inclinations allows us to test whether the sample is inconsistent with the expected inclination angles of a sample of randomly oriented sample of objects. This may be expected since equator-on objects tend to show higher variability amplitudes \citep{Vos2017} so their rotation periods are more easily detected and measured with variability observations. {Moreover, brown dwarfs that are inclined pole-on should not exhibit variability due to rotational modulation of atmospheric inhomogeneities, and this is evident by our lack of detections in brown dwarfs with inclinations $<20^{\circ}$ (Figure \ref{fig:inclination_amplitude}). Thus we would expect that the sample of variable objects should be biased towards equator-on objects.} For a sample of objects with random orientation, the probability distribution of inclination angles is $P(i)\sim \sin(i)$ \citep{Vos2017}. We perform a 1-D Kolmogorov-Smirnov test to determine if the measured inclination angles differ significantly from a randomly oriented distribution of inclination angles. We find a Kolmogorov-Smirnov test statistic of $\sim0.18$, and a significance level of $0.5$. Thus there is currently no evidence that the sample of measured inclinations differs from that expected from randomly oriented objects. This will be a useful test to run as a larger sample of inclination angles are measured in the future.}

\begin{table*}[]
\centering
\begin{tabular}{llllll}
\hline \hline
Target       & $v \sin(i)$ (km/s)        & RV (km/s)   & Period (hr)  & Radius Estimate  & Inclination \\ [0.5ex]  \hline
\obj{2m0045} &  $31.76^{+0.45}_{-0.41}$  &$5.19^{+0.22}_{-0.25}$   &   $2.4\pm0.1$      &$1.62\pm0.06$     & $22\pm1^{\circ}$    \\
\obj{2m0501} &  $9.57^{+0.67}_{-0.58}$   &$ 24.65^{+0.14}_{-0.17}$   &   $15.7\pm0.2$ &$1.38\pm0.18$  &${60^{+20 }_{-9}} ^{\circ}$ \\
\obj{2m1425} & $33.08^{+0.53}_{-0.49}$   &$ 5.38\pm0.27$   &     $2.5-5.6$           & $1.32\pm0.09$         & ${54^{+36}_{-15}}^{\circ}$       \\ [0.5ex]  \hline  
\end{tabular}
\caption{Measured $v\sin (i)$ values, rotation periods and inclinations for \obj{2m0045}, \obj{2m0501} and \obj{2m1425}.}
\label{tab:inclination}
\end{table*}

\subsection{The variability amplitude is influenced by the inclination angle}
The addition of three more objects with measured variability in the mid-IR and measured inclination allows us to further test the relationships between inclination, variability amplitude and color anomaly introduced by \cite{Vos2017}. In the left panel of Figure \ref{fig:inclination_amplitude} we update the plot showing the tentative relation between inclination angle and [3.6] variability amplitude and in the right panel of Figure \ref{fig:inclination_amplitude} we present the equivalent plot for [4.5] variability data. Both plots show that the population of high-amplitude variables are viewed close to equator-on ($90^{\circ}$), while the maximum variability amplitudes decrease as the object is viewed closer to pole-on ($0^{\circ}$). 

We test the significance of the inclination dependence of the variability amplitude using the two-dimensional two-sample Kolmogorov-Smirnov test \citep{Peacock1983, Fasano1987}, which is used to assess the null hypothesis that two samples are drawn from the same two-dimensional distribution. We simulate a random distribution of inclinations and amplitudes with the same sample size and ranges as our data in Figure \ref{fig:inclination_amplitude}. The two-dimensional Kolmogorov-Smirnov test outputs a test statistic which is a measure of the cumulative probability difference between the two distributions, and the $p$-value, the probability that the two samples are drawn from the same distribution -- in our case this is the probability that the data in Figure \ref{fig:inclination_amplitude} are drawn from a random sample. We run the two-dimensional Kolmogorov-Smirnov test on the $3.6~\mu$m and $4.5~\mu$m data 1000 times each to obtain a distribution of the test statistic and $p$-value in each case. For the $3.6~\mu$m data, we find a Kolmogorov-Smirnov test statistic of 0.8 with a $p$-value of $2\times10^{-5}$, indicating that the data is significantly different from a random sample of inclinations and amplitudes. Similarly, the $4.5~\mu$m data gives a test statistic of 0.8 with a $p$-value of $4\times10^{-5}$. Thus the effect of inclination angle on the measured variability amplitude appears to be highly significant. 
The large body of variability studies carried out to date has shown that the variability properties of brown dwarfs are affected by fundamental properties such as temperature \citep[e.g.][]{Radigan2014, Metchev2015a} and gravity \citep[e.g.][]{Biller2018, Vos2019}. In Figure \ref{fig:inclination_amplitude}, this is evident by the range of amplitudes that have been observed in objects that are viewed with inclination angles close to $90^{\circ}$. Our analysis shows that inclination is a secondary effect that reduces the intrinsic variability signal for objects not viewed equator-on.

\begin{figure}[tb]
   \centering
       \includegraphics[scale=0.8]{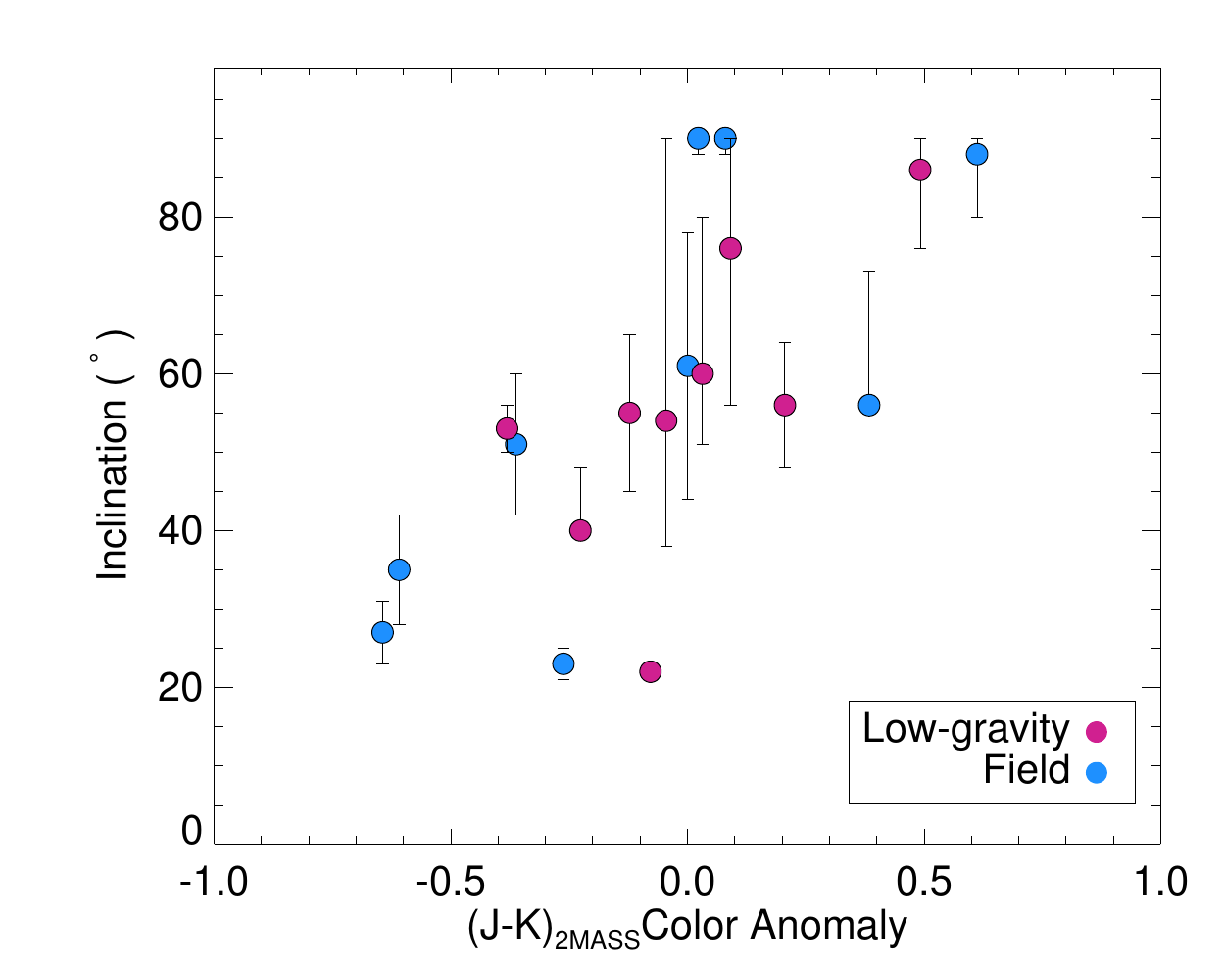}
   \caption{Inclination angle plotted against $(J-K)_{2MASS}$ color anomaly for variable young and field brown dwarfs.  The color anomaly of a brown dwarf is defined as median $(J-K)_{2MASS}$ color of brown dwarfs with the same spectral type and gravity class subtracted from the $(J-K)_{2MASS}$ color of the object. We find a tentative correlation between inclination angle and colour, where objects viewed equator-on appear redder than the median.}
   \label{fig:inclination_color}
\end{figure}

\subsection{The relation between inclination angle and color anomaly}
We additionally examine the relation between inclination and color anomaly in Figure \ref{fig:inclination_color}. The color anomaly of a brown dwarf is defined as median $(J-K)_{2MASS}$ color of brown dwarfs with the same spectral type and gravity class subtracted from the $(J-K)_{2MASS}$ color of the object \citep{Vos2017}. Objects with a positive color anomaly appear redder than the median and objects with a negative color anomaly appear bluer than the median. 
{Red colors in brown dwarfs are generally explained by the presence of thick clouds or hazes \citep{Marley2012, Gizis2012, Hiranaka2016}.}
Both the young and field dwarfs seem to follow the apparent trend between high-inclination and red colors in Figure \ref{fig:inclination_color}. 
Following \citet{Press1987}, we calculate the Spearman's $\rho$ coefficient to determine the significance of this apparent relation. 
For the field population, we find a correlation coefficient of 0.71 with a $2.1\sigma$ significance. For the young population, we find a correlation coefficient of 0.73 with a $2.1\sigma$ significance. For the combined population of both field and young object, we find a correlation coefficient of 0.76 with a $3.2\sigma$ significance. 
As discussed in \citet{Vos2017}, one explanation for this correlation would be the accumulation of thicker clouds at the equator relative to the clouds, similar to the zonally banded cloud patterns observed on Jupiter. Recently, \citet{Showman2019} presented global, three-dimensional atmospheric simulations of brown dwarfs and giant planets, showing that atmospheric waves and turbulence interact with the rotation to produce numerous zonal jets. While these models do not include clouds, it is possible that the clouds may couple with these jets to form a zonally banded cloud pattern, as we see on Jupiter \citep{Showman2019}. 


\section{Rotation Periods of Young Brown Dwarfs} \label{sec:periods}

\begin{figure*}[tb]
   \centering
   \includegraphics[scale=0.75]{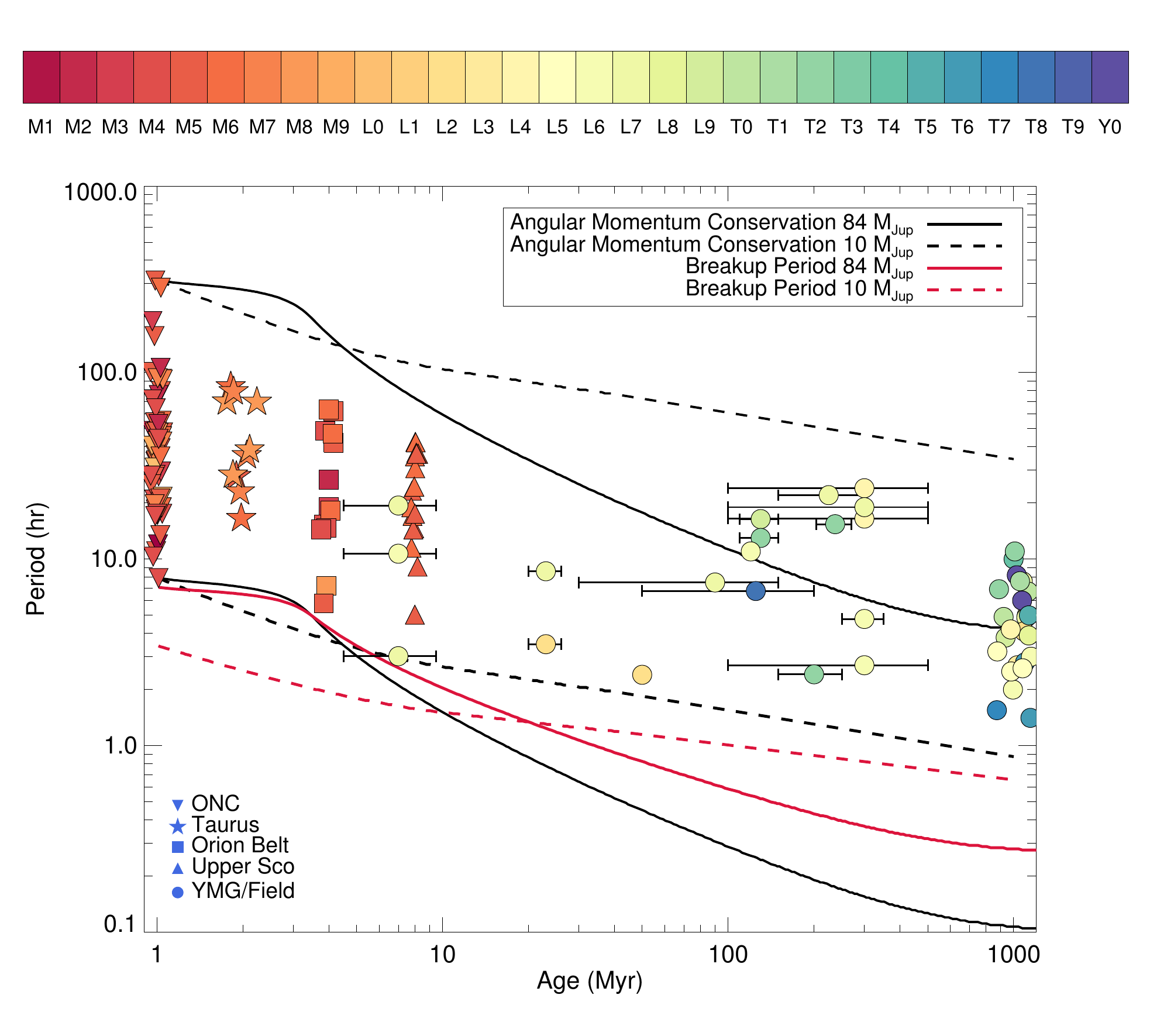}
   \caption{Rotation rates of brown dwarfs as a function of age. The ages of the field dwarf sample are not known, and we plot them at 1 Gyr. Black lines show the expected rotation period evolution of a $10~M_{\mathrm{Jup}}$ (dashed) and $84~M_{\mathrm{Jup}}$ (solid) object assuming conservation of angular momentum, with initial rotation periods as the maximum and minimum periods measured for the 1 Myr sample, and using evolutionary models \citet{Baraffe2015}. The breakup period for masses of $10~M_{\mathrm{Jup}}$ (dashed) and $84~M_{\mathrm{Jup}}$ (solid) are plotted in red. The population of brown dwarfs with measured rotation periods agree with the calculated periods using evolutionary models and assuming angular momentum conservation.}
   \label{fig:rotation_age}
\end{figure*}

We combine our two new rotation periods with the rich sample of variable brown dwarfs to investigate the relations between rotation period and age. \citet{Moore2019} provide a well-vetted sample of brown dwarfs with measured rotation periods with ages ($<10$ Myr). 
This young sample includes variable brown dwarfs in the 1 Myr Orion Nebula Cluster reported by \citet{Rodriguez-Ledesma2009}, 
members of the $3-6$ Myr Orion Belt region reported by \citet{Scholz2004a, Scholz2005, Cody2010}, members of Upper Sco \citep[$\sim$10 Myr,][]{Scholz2015, Moore2019} and members of Taurus \citep[$\sim 1$ Myr,][]{Scholz2018}. 
{\citet{Moore2019} define absolute $J$-band magnitude cutoffs to separate brown dwarfs from low-mass stars in this sample. In this $<10$ Myr sample, the lower mass limits are at $\sim0.02~M_{\odot}$ \citep{Moore2019}. We point the reader to this paper for a detailed discussion of the vetting procedure for the $<10$ Myr sample. }
The intermediate age ($10-500$ Myr) sample is compiled from variability monitoring programs reported by \citet{Metchev2015a, Zhou2016,Biller2018, Schneider2018, Vos2018, Zhou2018, Manjavacas2019}. The field age sample is composed of variable dwarfs reported by \citet{Metchev2015a, Yang2016, Esplin2016, Cushing2016, Leggett2016}. Figure \ref{fig:rotation_age} shows the full sample of brown dwarfs with measured rotation periods from 1 Myr to 1 Gyr. For the Orion Nebula Cluster, the Orion Belt region, Upper Sco and Taurus, we have added a random spread in the ages of each object for clarity. As pointed out by \citet{Moore2019}, age uncertainties and age spreads for young clusters, young moving groups and for the field sample are likely sources of error. 
As previously found by \citet{Scholz2015, Schneider2018, Moore2019}, we find that the sample of variable brown dwarfs spin-up over time, due in part to angular momentum conservation as the objects contract and possible disk-locking at very young age \citep[$<5$ Myr;][]{Scholz2018}.

We investigate whether the current sample of measured rotation periods in consistent with gravitational contraction using evolutionary models and assuming angular momentum conservation \citep{Bouvier2014, Scholz2015, Schneider2018}. We use the evolutionary models of \citet{Baraffe2015} to estimate the radii of brown dwarfs at 1 Myr with masses of $84~M_{\mathrm{Jup}}$ and $10~M_{\mathrm{Jup}}$ and the maximum and minimum measured periods from this age group. Assuming angular momentum conservation, we then calculated the expected periods from 1 Myr - 1 Gyr using the predicted radii from \citet{Baraffe2015}. Since the \citet{Baraffe2015} models do not reach 1 Gyr for the low-mass objects, we artificially add a final radius data point of $1~R_{\mathrm{Jup}}$ at 1 Gyr. This additional radius value is also added by \citet{Schneider2018}, but is not explained in the text. In Figure \ref{fig:rotation_age}, we plot the period evolution assuming angular moment conservation for two brown dwarfs in black.

The lower limit for the rotation period of a rotating object is known as the ``breakup period''. We calculate the breakup period by equating the equatorial velocity with the escape velocity \citep{Leggett2016}, which changes in time as the radius contracts. {This method assumes a spherical object, however in reality a rapidly rotating object will experience strong flattening.} In Figure \ref{fig:rotation_age} we plot the breakup period of a $10~M_{\mathrm{Jup}}$ and $84~M_{\mathrm{Jup}}$ object in red. 
The brown dwarfs in Figure \ref{fig:rotation_age} show a wide spread in rotation period measurements at each age, but the ``spin-up'' of brown dwarfs with age is evident. Comparing with the angular momentum conservation and break-up limits based on evolutionary models, we see that the general population of brown dwarfs with measured rotation periods agree with these limits.
{Additional variability data for members of young moving groups as part of our 30 object \textit{Spitzer} Program 14128 (PI: J. Faherty, Vos et al. in prep)} will allow us to fill in the intermediate age sample, and further investigate the angular momentum evolution of brown dwarfs over their lifetime, and in particular investigate the effects of mass on the spin-up process, as has been studied for the $<10~$Myr population by \citet{Scholz2018}. 

\section{Conclusions}
In this paper we present \textit{Spitzer} $3.6~\mu$m and $4.5~\mu$m variability monitoring observations of three young L dwarfs previously found to exhibit variability in the near-IR \citep{Vos2019}, \obj{2m0045}, \obj{2m0501} and \obj{2m1425}.

\begin{itemize}

\item  We use new and published high-resolution spectra to measure the rotational velocities of \obj{2m0045}, \obj{2m0501} and \obj{2m1425}. We combine these $v\sin(i)$ values with estimated radii to place upper limits on the rotation periods of each target, which guided our \textit{Spitzer} observation lengths.

\item We detect significant periodic variability in \obj{2m0045} and \obj{2m0501} at $3.6~\mu$m and $4.5~\mu$m, but do not detect variability in \obj{2m1425}.

\item We measure rotation periods of $2.4\pm0.1$ hr for \obj{2m0045} and $15.7\pm0.2$ hr for \obj{2m0501}.  Since \obj{2m1425} does not show variability in our \textit{Spitzer} observations, we do not measure a rotation period for this object. However, based on our earlier ground-based monitoring and measured $v\sin(i)$, we can estimate a rotation period of $2.4-5.6$ hr for \obj{2m1425}.

\item We combine these new \textit{Spitzer} detections with the growing sample of young and field brown dwarfs with detected mid-IR variability to investigate their variability amplitudes, inclination angles and rotation periods, particularly focusing on potential differences between the young and field brown dwarf populations. 

\item Previous variability searches found correlations between low-gravity and high-amplitude variability for L dwarfs, both in the mid-IR \citep{Metchev2015a} and near-IR \citep{Vos2019}. We find no evidence for an increase in $3.6~\mu$m amplitude in the young dwarfs compared to the field population. At $4.5~\mu$m, we do not find an amplitude enhancement for young, early-L dwarfs. However, there is an apparent enhancement of $4.5~\mu$m amplitude for young, late-L dwarfs.

\item We  calculate the inclination angles of each target, finding inclinations of $22\pm1^{\circ}$ for \obj{2m0045}, ${60^{+20 }_{-9}} ^{\circ}$ for \obj{2m0501} and ${54^{+36}_{-16}}^{\circ}$ for \obj{2m1425}. These new inclination angles are consistent with the tentative relations between inclination, amplitude and colour reported by \citet{Vos2017}. The largest variability amplitudes are observed for brown dwarfs that are observed equator-on, and the maximum amplitude decreases for lower inclinations. 

\item We find a correlation between inclination angle and $(J-K)_{\mathrm{2MASS}}$ for the sample of brown dwarfs with measured inclinations. This suggests that brown dwarfs viewed equator-on appear redder than the median and brown dwarfs viewed pole-on appear bluer than the median.

\item Finally, we compile the rotation periods of a large sample of brown dwarfs with ages 1 Myr to field ages and compare the rotation rates predicted by evolutionary models assuming angular momentum conservation. We find that the rotation rates of the current sample of brown dwarfs generally falls within the limits set by evolutionary models and breakup limits. In future work we will examine the mass dependence of the spin-up process.
\end{itemize}

\section{Acknowledgements}
The authors would like to thank the anonymous referee and scientific editor for extensive feedback that significantly improved the clarity of the paper. 
The authors would like to thank Adam Schneider, Mark Marley and Alex Scholz for helpful discussions regarding the angular momentum evolution of brown dwarfs, and Jack Gallimore for his contributions to the DREAM-ZS fitting code. 
This work is based on observations made with the Spitzer Space Telescope, which is operated by the Jet Propulsion Laboratory, California Institute of Technology under a contract with NASA. Based on observations obtained at the Gemini Observatory, which is operated by the Association of Universities for Research in Astronomy, Inc., under a cooperative agreement with the NSF on behalf of the Gemini partnership: the National Science Foundation (United States), National Research Council (Canada), CONICYT (Chile), Ministerio de Ciencia, Tecnolog\'{i}a e Innovaci\'{o}n Productiva (Argentina), Minist\'{e}rio da Ci\^{e}ncia, Tecnologia e Inova\c{c}\~{a}o (Brazil), and Korea Astronomy and Space Science Institute (Republic of Korea).
JV acknowledges support by NSF Award Number 1614527 and \textit{Spitzer} Cycle 14 JPL Research Support Agreement 1627378. 
T.J.D.\ is supported by the international Gemini Observatory, a program of NSF’s OIR Lab, which is managed by the Association of Universities for Research in Astronomy (AURA) under a cooperative agreement with the National Science Foundation, on behalf of the Gemini partnership of Argentina, Brazil, Canada, Chile, the Republic of Korea, and the United States of America.
T.H. acknowledges support from the European Research Council under the Horizon 2020 Framework Program via the ERC Advanced Grant Origins 83 24 28. The authors wish to recognize and acknowledge the very significant cultural role and reverence that the summit of Mauna Kea has always had within the indigenous Hawaiian community. We are most fortunate to have the opportunity to conduct observations from this mountain.

\bibliography{Full_paper.bib}

\begin{thebibliography}{}
\expandafter\ifx\csname natexlab\endcsname\relax\def\natexlab#1{#1}\fi
\providecommand{\url}[1]{\href{#1}{#1}}

\bibitem[{Allard {et~al.}(2012)Allard, Homeier, \& Freytag}]{Allard2012}
Allard, F., Homeier, D., \& Freytag, B. 2012, Philosophical Transactions of the
  Royal Society A: Mathematical, Physical and Engineering Sciences, 370, 2765

\bibitem[{Allers {et~al.}(2016)Allers, Gallimore, Liu, \& Dupuy}]{Allers2016}
Allers, K.~N., Gallimore, J.~F., Liu, M.~C., \& Dupuy, T.~J. 2016, \apj, 819,
  133

\bibitem[{Allers \& Liu(2013)}]{Allers2013}
Allers, K.~N., \& Liu, M.~C. 2013, \apj, 772, 79

\bibitem[{{Allers} {et~al.}(2020){Allers}, {Vos}, {Biller}, \&
  {Williams}}]{Allers2020}
{Allers}, K.~N., {Vos}, J.~M., {Biller}, B.~A., \& {Williams}, P. K.~G. 2020,
  Science, 368, 169

\bibitem[{Apai {et~al.}(2016)Apai, Kasper, Skemer, Hanson, Lagrange, Biller,
  Bonnefoy, Buenzli, \& Vigan}]{Apai2016}
Apai, D., Kasper, M., Skemer, A., {et~al.} 2016, \apj, 820, 40

\bibitem[{Apai {et~al.}(2017)Apai, Karalidi, Marley, Yang, Flateau, Metchev,
  Cowan, Buenzli, Burgasser, Radigan, Artigau, \& Lowrance}]{Apai2017}
Apai, D., Karalidi, T., Marley, M.~S., {et~al.} 2017, Science, 357, 683

\bibitem[{Artigau(2018)}]{Artigau2018}
Artigau, {\'{E}}. 2018, in Handbook of Exoplanets (Cham: Springer International
  Publishing), 555--573

\bibitem[{Baraffe {et~al.}(2015)Baraffe, Homeier, Allard, \&
  Chabrier}]{Baraffe2015}
Baraffe, I., Homeier, D., Allard, F., \& Chabrier, G. 2015, \aap, 577, 4

\bibitem[{Best {et~al.}(2017)Best, Liu, Dupuy, \& Magnier}]{Best2017}
Best, W. M.~J., Liu, M.~C., Dupuy, T.~J., \& Magnier, E.~A. 2017, 3, 1706.01883

\bibitem[{Biller(2017)}]{Biller2017}
Biller, B. 2017, Astronomical Review, 2857, 1

\bibitem[{Biller {et~al.}(2015)Biller, Vos, Bonavita, Buenzli, Baxter,
  Crossfield, Allers, Liu, Bonnefoy, Deacon, Brandner, Schlieder, Dupuy,
  Kopytova, Manjavacas, Allard, Homeier, \& Henning}]{Biller2015}
Biller, B.~A., Vos, J., Bonavita, M., {et~al.} 2015, \apjl, 813, 1

\bibitem[{Biller {et~al.}(2018)Biller, Vos, Buenzli, Allers, Bonnefoy, Charnay,
  B{\'{e}}zard, Allard, Homeier, Bonavita, Brandner, Crossfield, Dupuy,
  Henning, Kopytova, Liu, Manjavacas, \& Schlieder}]{Biller2018}
Biller, B.~A., Vos, J., Buenzli, E., {et~al.} 2018, \aj, 155, 95

\bibitem[{Blake {et~al.}(2010)Blake, Charbonneau, \& White}]{Blake2010}
Blake, C.~H., Charbonneau, D., \& White, R.~J. 2010, \apj, 723, 684

\bibitem[{{Bouchy} {et~al.}(2010){Bouchy}, {Hebb}, {Skillen}, {Collier
  Cameron}, {Smalley}, {Udry}, {Anderson}, {Boisse}, {Enoch}, {Haswell},
  {H{\'e}brard}, {Hellier}, {Joshi}, {Kane}, {Maxted}, {Mayor}, {Moutou},
  {Pepe}, {Pollacco}, {Queloz}, {S{\'e}gransan}, {Simpson}, {Smith},
  {Stempels}, {Street}, {Triaud}, {West}, \& {Wheatley}}]{Bouchy2010}
{Bouchy}, F., {Hebb}, L., {Skillen}, I., {et~al.} 2010, \aap, 519, A98

\bibitem[{Bouvier {et~al.}(2014)Bouvier, Matt, Mohanty, Scholz, Stassun, \&
  Zanni}]{Bouvier2014}
Bouvier, J., Matt, S.~P., Mohanty, S., {et~al.} 2014, in Protostars and Planets
  VI (University of Arizona Press)

\bibitem[{{Bowler} {et~al.}(2020){Bowler}, {Zhou}, {Morley}, {Kataria},
  {Bryan}, {Benneke}, \& {Batygin}}]{Bowler2020}
{Bowler}, B.~P., {Zhou}, Y., {Morley}, C.~V., {et~al.} 2020, \apjl, 893, L30

\bibitem[{Buenzli {et~al.}(2014)Buenzli, Apai, Radigan, Reid, \&
  Flateau}]{Buenzli2014}
Buenzli, E., Apai, D., Radigan, J., Reid, I.~N., \& Flateau, D. 2014, \apj,
  782, 77

\bibitem[{{Buenzli} {et~al.}(2015){Buenzli}, {Saumon}, {Marley}, {Apai},
  {Radigan}, {Bedin}, {Reid}, \& {Morley}}]{Buenzli2015a}
{Buenzli}, E., {Saumon}, D., {Marley}, M.~S., {et~al.} 2015, \apj, 798, 127

\bibitem[{{Buenzli} {et~al.}(2012){Buenzli}, {Apai}, {Morley}, {Flateau},
  {Showman}, {Burrows}, {Marley}, {Lewis}, \& {Reid}}]{Buenzli2012}
{Buenzli}, E., {Apai}, D., {Morley}, C.~V., {et~al.} 2012, \apjl, 760, L31

\bibitem[{Chabrier {et~al.}(2000)Chabrier, Baraffe, Allard, Hauschildt3, \&
  P}]{Chabrier2000}
Chabrier, G., Baraffe, I., Allard, F., Hauschildt3, \& P. 2000, \apj, 542, 464

\bibitem[{Clarke {et~al.}(2008)Clarke, Hodgkin, Oppenheimer, Robertson, \&
  Haubois}]{Clarke2008}
Clarke, F.~J., Hodgkin, S.~T., Oppenheimer, B.~R., Robertson, J., \& Haubois,
  X. 2008, \mnras, 386, 2009

\bibitem[{{Cody} \& {Hillenbrand}(2010)}]{Cody2010}
{Cody}, A.~M., \& {Hillenbrand}, L.~A. 2010, \apjs, 191, 389

\bibitem[{{Croll} {et~al.}(2016){Croll}, {Muirhead}, {Han}, {Dalba}, {Radigan},
  {Morley}, {Lazarevic}, \& {Taylor}}]{Croll2016}
{Croll}, B., {Muirhead}, P.~S., {Han}, E., {et~al.} 2016, arXiv e-prints,
  arXiv:1609.03586

\bibitem[{Cruz {et~al.}(2009)Cruz, Kirkpatrick, \& Burgasser}]{Cruz2009}
Cruz, K.~L., Kirkpatrick, J.~D., \& Burgasser, A.~J. 2009, \aj, 137, 3345

\bibitem[{Cushing {et~al.}(2016)Cushing, Hardegree-Ullman, Trucks, Morley,
  Gizis, Marley, Fortney, Kirkpatrick, Gelino, Mace, \& Carey}]{Cushing2016}
Cushing, M.~C., Hardegree-Ullman, K.~K., Trucks, J.~L., {et~al.} 2016, \apj,
  823, 152

\bibitem[{{Deleuil} {et~al.}(2008){Deleuil}, {Deeg}, {Alonso}, {Bouchy},
  {Rouan}, {Auvergne}, {Baglin}, {Aigrain}, {Almenara}, {Barbieri}, {Barge},
  {Bruntt}, {Bord{\'e}}, {Collier Cameron}, {Csizmadia}, {de La Reza},
  {Dvorak}, {Erikson}, {Fridlund}, {Gandolfi}, {Gillon}, {Guenther}, {Guillot},
  {Hatzes}, {H{\'e}brard}, {Jorda}, {Lammer}, {L{\'e}ger}, {Llebaria},
  {Loeillet}, {Mayor}, {Mazeh}, {Moutou}, {Ollivier}, {P{\"a}tzold}, {Pont},
  {Queloz}, {Rauer}, {Schneider}, {Shporer}, {Wuchterl}, \&
  {Zucker}}]{Deleuil2008}
{Deleuil}, M., {Deeg}, H.~J., {Alonso}, R., {et~al.} 2008, \aap, 491, 889

\bibitem[{{Delrez} {et~al.}(2018){Delrez}, {Gillon}, {Triaud}, {Demory}, {de
  Wit}, {Ingalls}, {Agol}, {Bolmont}, {Burdanov}, {Burgasser}, {Carey},
  {Jehin}, {Leconte}, {Lederer}, {Queloz}, {Selsis}, \& {Van
  Grootel}}]{Delrez2018}
{Delrez}, L., {Gillon}, M., {Triaud}, A.~H.~M.~J., {et~al.} 2018, \mnras, 475,
  3577

\bibitem[{Dupuy \& Liu(2012)}]{Dupuy2012}
Dupuy, T.~J., \& Liu, M.~C. 2012, \apjs, 201, 19

\bibitem[{Eriksson {et~al.}(2019)Eriksson, Janson, \&
  Calissendorff}]{Eriksson2019}
Eriksson, S., Janson, M., \& Calissendorff, P. 2019, Astronomy {\&}
  Astrophysics, 145, 1

\bibitem[{Esplin {et~al.}(2016)Esplin, Luhman, Cushing, Hardegree-Ullman,
  Trucks, Burgasser, \& Schneider}]{Esplin2016}
Esplin, T.~L., Luhman, K.~L., Cushing, M.~C., {et~al.} 2016, \apj, 832, 58

\bibitem[{Faherty {et~al.}(2016)Faherty, Riedel, Cruz, Gagne, Filippazzo,
  Lambrides, Fica, Weinberger, Thorstensen, Tinney, Baldassare, Lemonier, \&
  Rice}]{Faherty2016}
Faherty, J.~K., Riedel, A.~R., Cruz, K.~L., {et~al.} 2016, \apjs, 225, 1

\bibitem[{{Fasano} \& {Franceschini}(1987)}]{Fasano1987}
{Fasano}, G., \& {Franceschini}, A. 1987, \mnras, 225, 155

\bibitem[{Fazio {et~al.}(2004)Fazio, Hora, Allen, Ashby, Barmby, Deutsch,
  Huang, Kleiner, Marengo, Megeath, Melnick, Pahre, Patten, Polizotti, Smith,
  Taylor, Wang, Willner, Hoffmann, Pipher, Forrest, McMurty, McCreight,
  McKelvey, McMurray, Koch, Moseley, Arendt, Mentzell, Marx, Losch, Mayman,
  Eichhorn, Krebs, Jhabvala, Gezari, Fixsen, Flores, Shakoorzadeh, Jungo,
  Hakun, Workman, Karpati, Kichak, Whitley, Mann, Tollestrup, Eisenhardt,
  Stern, Gorjian, Bhattacharya, Carey, Nelson, Glaccum, Lacy, Lowrance, Laine,
  Reach, Stauffer, Surace, Wilson, Wright, Hoffman, Domingo, \&
  Cohen}]{Fazio2004}
Fazio, G.~G., Hora, J.~L., Allen, L.~E., {et~al.} 2004, \apjs, 154, 10

\bibitem[{Filippazzo {et~al.}(2015)Filippazzo, Rice, Faherty, Cruz, {Van
  Gordon}, \& Looper}]{Filippazzo2015}
Filippazzo, J.~C., Rice, E.~L., Faherty, J., {et~al.} 2015, \apj, 810, 158

\bibitem[{Foreman-Mackey {et~al.}(2013)Foreman-Mackey, Hogg, Lang, \&
  Goodman}]{fm2013}
Foreman-Mackey, D., Hogg, D.~W., Lang, D., \& Goodman, J. 2013, \pasp, 125, 306

\bibitem[{Gagn{\'{e}} {et~al.}(2015)Gagn{\'{e}}, Faherty, Cruz,
  Lafreni{\'{e}}re, Doyon, Malo, Burgasser, Naud, Artigau, Bouchard, Gizis, \&
  Albert}]{Gagne2015c}
Gagn{\'{e}}, J., Faherty, J.~K., Cruz, K.~L., {et~al.} 2015, \apjs, 219, 33

\bibitem[{Gagn{\'{e}} {et~al.}(2017)Gagn{\'{e}}, Faherty, Burgasser, Artigau,
  Bouchard, Albert, Lafreni{\`{e}}re, Doyon, \& Gagliuffi}]{Gagne2017}
Gagn{\'{e}}, J., Faherty, J.~K., Burgasser, A.~J., {et~al.} 2017, \apj, 841, L1

\bibitem[{Gillon {et~al.}(2013)Gillon, Triaud, Jehin, Delrez, Opitom, Magain,
  Lendl, \& Queloz}]{Gillon2013}
Gillon, M., Triaud, a. H. M.~J., Jehin, E., {et~al.} 2013, \aap, 555, L5

\bibitem[{Girardin {et~al.}(2013)Girardin, Artigau, \& Doyon}]{Girardin2013}
Girardin, F., Artigau, {\'{E}}., \& Doyon, R. 2013, \apj, 767, 61

\bibitem[{Gizis {et~al.}(2012)Gizis, Faherty, Liu, Castro, Shaw, Vrba, Harris,
  Aller, \& Deacon}]{Gizis2012}
Gizis, J.~E., Faherty, J.~K., Liu, M.~C., {et~al.} 2012, \aj, 144, 94

\bibitem[{Heinze {et~al.}(2015)Heinze, Metchev, \& Kellogg}]{Heinze2015}
Heinze, A.~N., Metchev, S., \& Kellogg, K. 2015, \apj, 801, arXiv:1412.6733

\bibitem[{Heinze {et~al.}(2013)Heinze, Metchev, Apai, Flateau, Kurtev, Marley,
  Radigan, Burgasser, Artigau, \& Plavchan}]{Heinze2014}
Heinze, A.~N., Metchev, S., Apai, D., {et~al.} 2013, \apj, 767, 173

\bibitem[{Hiranaka {et~al.}(2016)Hiranaka, Cruz, Douglas, Marley, \&
  Baldassare}]{Hiranaka2016}
Hiranaka, K., Cruz, K.~L., Douglas, S.~T., Marley, M.~S., \& Baldassare, V.~F.
  2016, \apj, 830, 96

\bibitem[{Knutson {et~al.}(2008)Knutson, Charbonneau, Allen, Burrows, \&
  Megeath}]{Knutson2008}
Knutson, H.~A., Charbonneau, D., Allen, L.~E., Burrows, A., \& Megeath, S.~T.
  2008, \apj, 673, 526

\bibitem[{Leggett {et~al.}(2016)Leggett, Cushing, Hardegree-Ullman, Trucks,
  Marley, Morley, Saumon, Carey, Fortney, Gelino, Gizis, Kirkpatrick, \&
  Mace}]{Leggett2016}
Leggett, S.~K., Cushing, M.~C., Hardegree-Ullman, K.~K., {et~al.} 2016, \apj,
  830, 141

\bibitem[{Lew {et~al.}(2016)Lew, Apai, Zhou, Schneider, Burgasser, Karalidi,
  Yang, Marley, Cowan, Bedin, Metchev, Radigan, \& Lowrance}]{Lew2016}
Lew, B. W.~P., Apai, D., Zhou, Y., {et~al.} 2016, \apj, 829, L32

\bibitem[{{Littlefair} {et~al.}(2014){Littlefair}, {Casewell}, {Parsons},
  {Dhillon}, {Marsh}, {G{\"a}nsicke}, {Bloemen}, {Catalan}, {Irawati}, {Hardy},
  {Mcallister}, {Bours}, {Richichi}, {Burleigh}, {Burningham}, {Breedt}, \&
  {Kerry}}]{Littlefair2014}
{Littlefair}, S.~P., {Casewell}, S.~L., {Parsons}, S.~G., {et~al.} 2014,
  \mnras, 445, 2106

\bibitem[{Liu {et~al.}(2016)Liu, Dupuy, \& Allers}]{Liu2016}
Liu, M.~C., Dupuy, T.~J., \& Allers, K.~N. 2016, \apj, 833, 96

\bibitem[{Liu {et~al.}(2013)Liu, Magnier, Deacon, Allers, Dupuy, Kotson, Aller,
  Burgett, Chambers, Draper, Hodapp, Jedicke, Kudritzki, Metcalfe, Morgan,
  Kaiser, Price, Tonry, \& Wainscoat}]{Liu2013}
Liu, M.~C., Magnier, E.~A., Deacon, N.~R., {et~al.} 2013, \apj, 777, L20

\bibitem[{Manjavacas {et~al.}(2018)Manjavacas, Apai, Zhou, Karalidi, Lew,
  Schneider, Cowan, Metchev, Miles-p{\'{a}}ez, Burgasser, Radigan, Bedin,
  Lowrance, \& Marley}]{Manjavacas2018}
Manjavacas, E., Apai, D., Zhou, Y., {et~al.} 2018, \aj, 155, 11

\bibitem[{Manjavacas {et~al.}(2019)Manjavacas, Apai, Lew, Zhou, Schneider,
  Burgasser, Karalidi, Miles-P{\'{a}}ez, Lowrance, Cowan, Bedin, Marley,
  Metchev, \& Radigan}]{Manjavacas2019}
Manjavacas, E., Apai, D., Lew, B. W.~P., {et~al.} 2019, \apj, 875, L15

\bibitem[{Marley {et~al.}(2012)Marley, Saumon, Cushing, Ackerman, Fortney, \&
  Freedman}]{Marley2012}
Marley, M.~S., Saumon, D., Cushing, M., {et~al.} 2012, \apj, 754, 135

\bibitem[{Metchev {et~al.}(2015)Metchev, Heinze, Apai, Flateau, Radigan,
  Burgasser, Marley, Artigau, Plavchan, \& Goldman}]{Metchev2015a}
Metchev, S.~A., Heinze, A., Apai, D., {et~al.} 2015, \apj, 799, 154

\bibitem[{{Miles-P{\'a}ez} {et~al.}(2019){Miles-P{\'a}ez}, {Metchev}, {Apai},
  {Zhou}, {Manjavacas}, {Karalidi}, {Lew}, {Burgasser}, {Bedin}, {Cowan},
  {Lowrance}, {Marley}, {Radigan}, \& {Schneider}}]{Miles-Paez2019}
{Miles-P{\'a}ez}, P.~A., {Metchev}, S., {Apai}, D., {et~al.} 2019, \apj, 883,
  181

\bibitem[{{Moore} {et~al.}(2019){Moore}, {Scholz}, \&
  {Jayawardhana}}]{Moore2019}
{Moore}, K., {Scholz}, A., \& {Jayawardhana}, R. 2019, \apj, 872, 159

\bibitem[{{Morales-Calder{\'o}n} {et~al.}(2006){Morales-Calder{\'o}n},
  {Stauffer}, {Kirkpatrick}, {Carey}, {Gelino}, {Barrado y Navascu{\'e}s},
  {Rebull}, {Lowrance}, {Marley}, {Charbonneau}, {Patten}, {Megeath}, \&
  {Buzasi}}]{Morales-Calderon2006}
{Morales-Calder{\'o}n}, M., {Stauffer}, J.~R., {Kirkpatrick}, J.~D., {et~al.}
  2006, \apj, 653, 1454

\bibitem[{Naud {et~al.}(2017)Naud, Artigau, Doyon, Malo, Gagn{\'{e}},
  Lafreni{\`{e}}re, Wolf, \& Magnier}]{Naud2017}
Naud, M.-e., Artigau, {\'{E}}., Doyon, R., {et~al.} 2017, \aj, 154, 129

\bibitem[{{Peacock}(1983)}]{Peacock1983}
{Peacock}, J.~A. 1983, \mnras, 202, 615

\bibitem[{{Pont} {et~al.}(2005){Pont}, {Melo}, {Bouchy}, {Udry}, {Queloz},
  {Mayor}, \& {Santos}}]{Pont2005}
{Pont}, F., {Melo}, C.~H.~F., {Bouchy}, F., {et~al.} 2005, \aap, 433, L21

\bibitem[{{Press} {et~al.}(1986){Press}, {Flannery}, \&
  {Teukolsky}}]{Press1987}
{Press}, W.~H., {Flannery}, B.~P., \& {Teukolsky}, S.~A. 1986, {Numerical
  recipes. The art of scientific computing}

\bibitem[{Radigan {et~al.}(2012)Radigan, Jayawardhana, Lafreni{\`{e}}re,
  Artigau, Marley, \& Saumon}]{Radigan2012}
Radigan, J., Jayawardhana, R., Lafreni{\`{e}}re, D., {et~al.} 2012, \apj, 750,
  105

\bibitem[{Radigan {et~al.}(2014)Radigan, Lafreni{\`{e}}re, Jayawardhana, \&
  Artigau}]{Radigan2014}
Radigan, J., Lafreni{\`{e}}re, D., Jayawardhana, R., \& Artigau, E. 2014, \apj,
  793, 75

\bibitem[{Riedel {et~al.}(2017)Riedel, Blunt, Lambrides, Rice, Cruz, \&
  Faherty}]{Riedel2017}
Riedel, A.~R., Blunt, S.~C., Lambrides, E.~L., {et~al.} 2017, \aj, 153, 1

\bibitem[{Riedel {et~al.}(2019)Riedel, DiTomasso, Rice, Alam, Abrahams, Crook,
  Cruz, \& Faherty}]{Riedel2019}
Riedel, A.~R., DiTomasso, V., Rice, E.~L., {et~al.} 2019

\bibitem[{Rodr{\'{i}}guez-Ledesma {et~al.}(2009)Rodr{\'{i}}guez-Ledesma, Mundt,
  Eisl{\"{o}}ffel, \& Herbst}]{Rodriguez-Ledesma2009}
Rodr{\'{i}}guez-Ledesma, M.~V., Mundt, R., Eisl{\"{o}}ffel, J., \& Herbst, W.
  2009, AIP Conference Proceedings, 1094, 118

\bibitem[{Saumon \& Marley(2008)}]{Saumon2008}
Saumon, D., \& Marley, M. 2008, \apj, 689, 1327

\bibitem[{Scargle(1982)}]{Scargle1982}
Scargle, J.~D. 1982, \apj, 263, 835

\bibitem[{{Schneider} {et~al.}(2018){Schneider}, {Hardegree-Ullman}, {Cushing},
  {Kirkpatrick}, \& {Shkolnik}}]{Schneider2018}
{Schneider}, A.~C., {Hardegree-Ullman}, K.~K., {Cushing}, M.~C., {Kirkpatrick},
  J.~D., \& {Shkolnik}, E.~L. 2018, \aj, 155, 238

\bibitem[{Scholz \& Eisl{\"{o}}ffel(2004)}]{Scholz2004a}
Scholz, A., \& Eisl{\"{o}}ffel, J. 2004, Astronomy and Astrophysics, 419, 249

\bibitem[{{Scholz} \& {Eisl{\"o}ffel}(2005)}]{Scholz2005}
{Scholz}, A., \& {Eisl{\"o}ffel}, J. 2005, \aap, 429, 1007

\bibitem[{Scholz {et~al.}(2015)Scholz, Kostov, Jayawardhana, \&
  Mu{\v{z}}i{\'{c}}}]{Scholz2015}
Scholz, A., Kostov, V., Jayawardhana, R., \& Mu{\v{z}}i{\'{c}}, K. 2015, \apj,
  809, L29

\bibitem[{Scholz {et~al.}(2018)Scholz, Moore, Jayawardhana, Aigrain, Peterson,
  \& Stelzer}]{Scholz2018}
Scholz, A., Moore, K., Jayawardhana, R., {et~al.} 2018, \apj, 859, 153

\bibitem[{{Schwarz}(1978)}]{Schwarz1978}
{Schwarz}, G. 1978, Annals of Statistics, 6, 461

\bibitem[{Showman {et~al.}(2019)Showman, Tan, \& Zhang}]{Showman2019}
Showman, A.~P., Tan, X., \& Zhang, X. 2019, \apj, 883, 4

\bibitem[{{Siverd} {et~al.}(2012){Siverd}, {Beatty}, {Pepper}, {Eastman},
  {Collins}, {Bieryla}, {Latham}, {Buchhave}, {Jensen}, {Crepp}, {Street},
  {Stassun}, {Gaudi}, {Berlind}, {Calkins}, {DePoy}, {Esquerdo}, {Fulton},
  {F{\H{u}}r{\'e}sz}, {Geary}, {Gould}, {Hebb}, {Kielkopf}, {Marshall},
  {Pogge}, {Stanek}, {Stefanik}, {Szentgyorgyi}, {Trueblood}, {Trueblood},
  {Stutz}, \& {van Saders}}]{Siverd2012}
{Siverd}, R.~J., {Beatty}, T.~G., {Pepper}, J., {et~al.} 2012, \apj, 761, 123

\bibitem[{Tremblin {et~al.}(2016)Tremblin, Amundsen, Chabrier, Baraffe,
  Drummond, Hinkley, Mourier, \& Venot}]{Tremblin2016}
Tremblin, P., Amundsen, D.~S., Chabrier, G., {et~al.} 2016, \apj, 817, L19

\bibitem[{Vos {et~al.}(2017)Vos, Allers, \& Biller}]{Vos2017}
Vos, J.~M., Allers, K.~N., \& Biller, B.~A. 2017, \apj, 842, 78

\bibitem[{Vos {et~al.}(2018)Vos, Allers, Biller, Liu, Dupuy, Gallimore,
  Adenuga, \& Best}]{Vos2018}
Vos, J.~M., Allers, K.~N., Biller, B.~A., {et~al.} 2018, \mnras, 474, 1041

\bibitem[{Vos {et~al.}(2019)Vos, Biller, Bonavita, Eriksson, Liu, Best,
  Metchev, Radigan, Allers, Janson, Buenzli, Dupuy, Bonnefoy, Manjavacas,
  Brandner, Crossfield, Deacon, Henning, Homeier, Kopytova, \&
  Schlieder}]{Vos2019}
Vos, J.~M., Biller, B.~A., Bonavita, M., {et~al.} 2019, \mnras, 483, 480

\bibitem[{Yang {et~al.}(2016)Yang, Apai, Marley, Karalidi, Flateau, Showman,
  Metchev, Buenzli, Radigan, Artigau, Lowrance, \& Burgasser}]{Yang2016}
Yang, H., Apai, D., Marley, M.~S., {et~al.} 2016, \apj, 826, 8

\bibitem[{{Zhou} {et~al.}(2016){Zhou}, {Apai}, {Schneider}, {Marley}, \&
  {Showman}}]{Zhou2016}
{Zhou}, Y., {Apai}, D., {Schneider}, G.~H., {Marley}, M.~S., \& {Showman},
  A.~P. 2016, \apj, 818, 176

\bibitem[{{Zhou} {et~al.}(2020){Zhou}, {Bowler}, {Morley}, {Apai}, {Kataria},
  {Bryan}, \& {Benneke}}]{Zhou2020}
{Zhou}, Y., {Bowler}, B.~P., {Morley}, C.~V., {et~al.} 2020, arXiv e-prints,
  arXiv:2004.05168

\bibitem[{Zhou {et~al.}(2018)Zhou, Apai, Metchev, Lew, Schneider, Marley,
  Karalidi, Manjavacas, Bedin, Cowan, Miles-P{\'{a}}ez, Lowrance, Radigan, \&
  Burgasser}]{Zhou2018}
Zhou, Y., Apai, D., Metchev, S., {et~al.} 2018, \aj, 155, 132

\bibitem[{{Zhou} {et~al.}(2019){Zhou}, {Apai}, {Lew}, {Schneider},
  {Manjavacas}, {Bedin}, {Cowan}, {Marley}, {Radigan}, {Karalidi}, {Lowrance},
  {Miles-P{\'a}ez}, {Metchev}, \& {Burgasser}}]{Zhou2019}
{Zhou}, Y., {Apai}, D., {Lew}, B. W.~P., {et~al.} 2019, \aj, 157, 128

\end{thebibliography}

\startlongtable
\begin{longrotatetable}
\begin{deluxetable}{lcccccccccc}
\tabletypesize{\scriptsize}
\tablecolumns{12}
\tablenum{3}
\tablewidth{0pt}
\setlength{\tabcolsep}{0.05in}
\tablecaption{Brown dwarf variability detections and upper limits in the infrared \label{tab:variables}}
\tablehead{  
\colhead{Target} &
\colhead{SpT} & 
\colhead{$A_{J}$} & 
\colhead{$A_{3.6}$} & 
\colhead{ $A_{4.5}$} & 
\colhead{Period } & 
\colhead{Age } & 
\colhead{Radius$^\mathrm{a}$} & 
\colhead{Companion} & 
\colhead{Inclination } & 
\colhead{Variability} \\
\colhead{} &
\colhead{} & 
\colhead{(\%)  } & 
\colhead{(\%) } & 
\colhead{(\%)} & 
\colhead{(hr)} & 
\colhead{(Myr)} & 
\colhead{($R_{\mathrm{Jup}}$)} & 
\colhead{} & 
\colhead{$(^{\circ})$} & 
\colhead{References}}
\startdata
2MASS J00132229-1143006	  &T2 (T3.5+T4.5?)	&$4.6\pm0.2$	&\nodata        &\nodata&	        $>2.8$             &	1000       &\nodata             &	0	        &\nodata                    &	1           \\
 LSPM J0036+1821          & L3.5         & $1.22\pm0.04$   & $0.47\pm0.05$   & $0.19\pm0.04$   & $2.7\pm0.3$          & 1000      & $1.01\pm0.07$       & 0   & $51\pm9$                       & 2,3,4        \\
2MASS J00452143+1634446   & L2      & $1.0\pm0.1$           & $0.18\pm0.04$   & $0.16\pm0.04$   & $2.4\pm0.1$          & 50        & $1.62\pm0.06$      & 0         & $23\pm1$            & 5,6         \\
2MASS J00470038+6803543   & L9        & 8                   & $1.07\pm0.04$   &                 & $16.4\pm0.2$        & $130\pm20$ & $1.28\pm0.02^8$      & 0         & $85^{+5}_{-9}$        & 7,4,8       \\
2MASS J00501994-3322402   & T7           & \nodata          & \textless{}0.59 & $1.07\pm0.11$   & $1.55\pm0.02$        & 1000       & $0.94\pm0.16$     & 0         & \nodata                  & 2            \\
2MASSI J0103320+193536    & L6          & \nodata         & $0.56\pm0.03$   & $0.87\pm0.09$     & $2.7\pm0.1$          & $300\pm200$ & $1.34\pm0.13$    & 0         & $40\pm8$               & 2            \\
2MASS J01075242+0041563   & L8           & \nodata          & $1.27\pm0.13$   & $1\pm0.2$       & $5\pm2$              & 1000        & $0.98\pm0.11$    & 0         & $56\pm17$               & 2            \\
GU PSC B                  & T3.5         & $4\pm1$         &  \nodata       &   \nodata         &  \nodata             & $130\pm20$  & \nodata          & 1         & \nodata                 & 9            \\
SIMP J013656.57+093347.3  & T2.5         & 4.5             & $1.5\pm0.2$     &  \nodata         & $2.414\pm0.078$     & $200\pm50$   & $1.22\pm0.01^12$    & 0         & $80\pm12$               & 10,11,12,13,4 \\
2MASS J01383648-0322181	  &T3	          &$5.5\pm1.2$	   &\nodata      	&\nodata	        &\nodata		       &1000	     &\nodata           &0	        &\nodata	               &2               \\
SDSS J015141.69+124429.6  & T0.0         & \nodata        & \textless{}0.83 & \textless{}0.81   & \nodata               & 1000       & $0.97\pm0.16$    & 0         & \nodata               & 2            \\
2MASSW J0310599+164816    & L8          &  \nodata         & \nodata         & \nodata           &   \nodata              & 1000     &\nodata          & 0         &  \nodata               & 14           \\
SDSS J042348.57-041403.5  & L6.5+T2      & $0.8\pm0.08$    &  \nodata          &  \nodata          & $2\pm0.4$            & 1000     &\nodata         & 0         & $79^{+11}_{-16}$          & 15,4         \\
PSO J071.8769-12.2713	  &T2	        &$4.5\pm0.6$	    &\nodata	        &\nodata	       &\nodata	        	&\nodata	 &\nodata           &0	        &\nodata	                &6      \\
2MASS J05012406-0010452   & L4           & $2\pm1$         & $0.36\pm0.04$   & $0.24\pm0.04$        & $15.7\pm0.2$    & $300\pm200$  & $1.38\pm0.18$    & 0         & $73^{+17}_{-12}$         & 5,6          \\
2MASS J05591914-1404488   & T4.5         & $0.7\pm0.5$     &   \nodata         & \nodata           & $10\pm3$           & 1000       & $0.97\pm0.11$      & 0         & \nodata                 & 10            \\
2MASS J06244595-4521548   & L6.5         & \textgreater{}1 &  \nodata          &  \nodata           &                    & 1000      & $0.99\pm0.10$    & 0         & \nodata                & 14           \\
SDSS J075840.33+324723.4  & T0.0+T3.5    & $4.8\pm0.2$     &  \nodata         &   \nodata         & $4.9\pm0.2$          & 1000      &\nodata           & 0         &  \nodata              & 10            \\
DENIS J081730.0-615520    & T6           & $0.6\pm0.1$     & \nodata          &  \nodata         & $2.8\pm0.2$          & 1000       & $0.94\pm0.16$    & 0         &  \nodata             & 10            \\
2MASSW J0820299+450031    & L5          &  \nodata        & \textless{}0.4  & \textless{}0.48       & \nodata               & 1000   &\nodata           & 0         &  \nodata        & 2            \\
2MASSI J0825196+211552    & L7.5        & \textgreater{}1 & $0.81\pm0.08$   & $1.4\pm0.3$           & $7.6\pm5$            & 1000    & $0.98\pm0.11$      & 0         &   \nodata       & 2            \\
SDSS J085834.42+325627.7  & T1           &  \nodata        & \textless{}0.27 & \textless{}0.64  &   \nodata             & 1000       &\nodata            & 0         &  \nodata           & 2            \\
2MASS J09490860-1545485   & T1.0+T2.0    & \nodata         & \textless{}0.54 & \textless{}0.83  &   \nodata            & 1000        & $0.96\pm0.17$     & 0         &   \nodata           & 2            \\
LP261-75B                 & L6V          & $2.4\pm0.14$    &   \nodata       & \nodata          & $4.78\pm0.98$       & $30\pm50$    &\nodata           & 1         &    \nodata            & 16           \\
2MASS J10101480-0406499	  &L6	        &$3.6\pm0.4$	    &\nodata	    &\nodata        	&\nodata	            &1000	     & $0.94\pm0.16$     &0	            &\nodata	        &11             \\
SDSS J104335.08+121314.1  & L9           &  \nodata        & $1.54\pm0.15$   & $1.2\pm0.2$      & $3.8\pm0.2$          & 1000        & $0.98\pm0.11$     & 0         &   \nodata           & 2            \\
WISE 1049B                & L7.5         &   \nodata       &  \nodata        & \nodata          & $4.87\pm0.01$        & 1000        & $1.02\pm0.07$     & 0         & $83^{+7}_{-8}$      & 17,18,4      \\
SDSS J105213.51+442255.7  & L6.5+T1.5    & $2.2\pm0.5$     & \nodata            &  \nodata         & $3\pm0.5$            & 1000     & \nodata           & 0         &  \nodata              & 19           \\
DENIS-P J1058.7-1548      & L2.5         & $0.8\pm0.1$ & $0.39\pm0.04$   & \textless{}0.3       & $4.1\pm0.2$          & 1000        & $1.00\pm0.07$     & 0         & $90^{+0}_{-2}$       & 20,2,4       \\
2MASS J10595185+3042059   & T8           &    \nodata       & \textless{}0.83 & \textless{}0.89 &    \nodata             & 1000      & \nodata           & 0         &  \nodata              & 2            \\
SDSS J111009.99+011613.0  & T5.5         & \nodata         &  \nodata       & \textless{}1.25 &  \nodata              & $130\pm20$   & $1.24\pm0.04$      & 0         & \nodata              & 8            \\
2MASS J11193254-1137466AB & L7           & \nodata          & $0.46\pm0.036$  & $0.96\pm0.037$  & $3.02\pm0.04$        & $7\pm2.5$   & \nodata            & 0         &  \nodata             & 21           \\
2MASS J11220826-3512363   & T2           &  \nodata        & \textless{}0.24 & \textless{}0.31 &                      & 1000         & \nodata            & 0         &  \nodata             & 2            \\
2MASS J11263991-5003550   & L5           & $1.2\pm0.1$     & $0.21\pm0.04$   & $0.29\pm0.15$   & $3.2\pm0.3$          & 1000         & \nodata            & 0         & $35\pm7$             & 10,2,4        \\
WISEA J114724.10-204021.3 & L7          &  \nodata         & $1.596\pm0.08$  & $2.216\pm0.09$  & $19.39\pm0.3$        & $7\pm2.5$    & \nodata          & 0         &   \nodata             & 21           \\
SDSS J115013.17+052012.3  & L6           &  \nodata          & \textless{}0.38 & \textless{}0.65 &                      & 1000       & \nodata            & 0         &  \nodata             & 2            \\
2M1207b                   & L6           & $2.72\pm0.1$            &  \nodata         &  \nodata         & $10.7\pm1$  & $7\pm2.5$   & $1.36\pm0.02$        & 1         &  \nodata        & 22           \\
2MASS J12095613-1004008   & T2+T7.5      &  \nodata         & \textless{}0.4  & \textless{}0.56 &  \nodata               & 1000      & \nodata          & 0         & \nodata               & 2            \\
2MASS J12195156+3128497   & L9           & \textgreater{}2 &   \nodata       & \nodata     &  \nodata              & 1000            & \nodata        & 0         & \nodata                 & 14           \\
SDSS J125453.90-012247.5  & T2         &    \nodata       & \textless{}0.15 & \textless{}0.3  &\nodata             & 1000            & $0.98\pm0.15$         & 0         & \nodata          & 2            \\
VHS1256-1257b             & L7         &    $24.7$       & \nodata           & $5.76\pm0.04$  &$22.04\pm0.05$      & 150-300         & \nodata            & 1         & $90^{+0}_{-29}$     & 23, 24            \\
Ross 458C                 & T8.5        & $2.62\pm0.02$   & \textless{}1.37 & \textless{}0.72 & $6.75\pm1.58$        & $125\pm75$    &\nodata         & 1         &   \nodata               & 25,2         \\
2MASS J13243559+6358284   & T2       &  \nodata         & $3.05\pm0.15$   & $3\pm0.3$       & $13\pm1$             & $130\pm20$      & \nodata       & 0         &\nodata                    & 2            \\
WISE J140518.39+553421.3  & Y0     &    \nodata        & $7.2\pm0.8$     & $7.1\pm0.2$   & $8.2\pm0.3$         & 1000                & \nodata     & 0         &   \nodata                 & 26           \\
ULAS J141623.94+134836.3  & T7.5        &   \nodata        & \textless{}0.91 & \textless{}0.59 &  \nodata              & 1000        & $0.96\pm0.16$         & 0         &  \nodata              & 2            \\
SDSS J141624.08+134826.7  & L6   &   \nodata        & \textless{}0.15 & \textless{}0.22 &   \nodata             & 1000               & \nodata      & 0         &  \nodata              & 2            \\
2MASS J14252798-3650229   & L4           & $0.7\pm0.3$     & \textless{}0.16 & \textless{}0.18 &  \nodata              & $130\pm20$  & $1.32\pm0.09$           & 0         & $52^{+19}_{-13}$      & 5,6          \\
2MASSW J1507476-162738    & L5          & 0.7             & $0.53\pm0.11$   & $0.45\pm0.09$   & $2.5\pm0.1$          & 1000          & $0.99\pm0.09$           & 0         & $23\pm2$              & 11,2,4       \\
SDSS J151114.66+060742.9  & L5.5+T5.0    &   \nodata        & $0.67\pm0.07$   & \textless{}0.49 & $11\pm2$             & 1000        & \nodata           & 0         & \nodata               & 2            \\
SDSS J151643.01+305344.4  & L8.0+L9.5    &  \nodata         & $2.4\pm0.2$     & $3.1\pm1.6$     & $6.7\pm5$            & 1000        & \nodata           & 0         &  \nodata              & 2            \\
SDSS J152039.82+354619.8  & T0           &  \nodata        & \textless{}0.3  & \textless{}0.45 &                      & 1000         & \nodata           & 0         &   \nodata             & 2            \\
SDSS J154508.93+355527.3  & L7.5         &   \nodata        & \textless{}0.59 & \textless{}1.15 &                     & 1000         & \nodata           & 0         &  \nodata              & 2           \\
2MASS J16154255+4953211   & L4           &   \nodata       & $0.9\pm0.2$     & \textless{}0.39 & $24\pm5$             & $300\pm200$  & $0.94\pm0.16$    & 0         & $86^{+4}_{-10}$       & 2,4          \\
2MASS J16291840+0335371   & T2           & $4.3\pm2.4$     & \nodata        &  \nodata    & $6.9\pm2.4$          & 1000              & \nodata            & 0         & $82^{+8}_{-12}$       & 10,4          \\
2MASSW J1632291+190441    & L8           &  \nodata         & $0.42\pm0.08$   & $0.5\pm0.3$     & $3.9\pm0.2$          & 1000        & $0.97\pm0.12$    & 0         &  \nodata              & 2            \\
2MASSI J1721039+334415    & L5.3       &    \nodata      & $0.33\pm0.07$   & \textless{}0.29 & $2.6\pm0.1$          & 1000           & \nodata            & 0         & $27\pm4$              & 2,4          \\
2MASSI J1726000+153819    & L2          &     \nodata      & \textless{}0.29 & \textless{}0.49 &                      & $300\pm200$  & $1.40\pm0.20$        & 0         &  \nodata              & 2            \\
WISEP J173835.52+273258.9 & Y0           &       \nodata   &    \nodata       & $3\pm0.1$       & $6\pm0.1$            & 1000        & \nodata            & 0         &   \nodata             & 27           \\
2MASS J17502484-0016151   & L5           & \textgreater{}1 &   \nodata       &  \nodata         &  \nodata               & 1000      & \nodata           & 0         & \nodata                & 14           \\
2MASS J17503293+1759042   & T3.5         & \textgreater{}1 &   \nodata      &   \nodata          & \nodata                & 1000     & $0.97\pm0.16$      & 0         &  \nodata              & 14           \\
2MASS J17534518-6559559   & L4           &    \nodata       & \textless{}0.25 &   \nodata       &   \nodata            & 1000        & \nodata           & 0         &   \nodata             & 2            \\
2MASS J18212815+1414010   & L5           &  \nodata         & $0.54\pm0.05$   & $0.71\pm0.14$   & $4.2\pm0.1$         & 1000         & \nodata           & 0         & $61\pm17$             & 11,2,4       \\
2MASS J18283572-4849046   & T5.5         & $0.9\pm0.1$     & \nodata          & \nodata           & $5\pm0.6$            & 1000      & $0.95\pm0.16$      & 0         & \nodata               & 10            \\
2MASS J20025073-0521524   & L5.5         & $1.7\pm0.2$     & \nodata           & \nodata           & \nodata        & $300\pm200$    & \nodata              & 0         & \nodata               & 5            \\
SDSS J204317.69-155103.4  & L9.5         & \nodata          & \textless{}0.71 & \textless{}0.74 &  \nodata             & 1000        & \nodata            & 0         & \nodata              & 2            \\
SDSS J205235.31-160929.8  & T1+T2.5 &                 & \textless{}0.36 & \textless{}0.71       &  \nodata            & 1000         & \nodata             & 0         &    \nodata          & 2            \\
PSO 318.5-22              & L7           & $10\pm1.3$      &                 & $3.4\pm0.08$    & $8.61\pm0.06$        & $23\pm3$     & $1.41\pm0.03$       & 0         & $56\pm8$             & 28,29,5      \\
2MASS J21392676+0220226   & L8.5+T3.5    & $26\pm1$            & $11\pm1$        & $10\pm1$        & $7.618\pm0.18$       & 1000     & $0.96\pm0.16$       & 0         & $90^{+0}_{-1}$        & 10,13,4,11    \\
HN PegB                   & T2.5         & 1.2             & $0.77\pm0.15$   & $1.1\pm0.5$     & $15.4\pm0.5$         & $237\pm33$   & \nodata          & 1         &   \nodata              & 2            \\
2MASS J21481628+4003593   & L7           & \nodata         & $1.33\pm0.07$   & $1.03\pm0.1$    & $19\pm4$             & $200\pm200$  & $0.99\pm0.10$           & 0         & $88^{+2}_{-8}$         & 2,4          \\
2MASSW J2208136+292121    & L2      &     \nodata      & $0.69\pm0.07$   & $0.54\pm0.11$   & $3.5\pm0.2$          & $23\pm3$         & $1.41\pm0.20$           & 0         & $55\pm10$             & 2,4          \\
2MASS J22153705+2110554	  &T1 (T0+T2?)	&$10.7\pm0.4$	&\nodata	        &\nodata	    &\nodata	        	&1000	         & \nodata           &0	        &\nodata	             &2               \\
2MASSW J2224438-015852    & L4.5V        &  \nodata      & \textless{}0.1  & \textless{}0.15 &                      & 1000           & $0.99\pm0.08$           & 0         &  \nodata              & 2            \\
2MASS J22282889-4310262   & T6.5         & 5.3             & $4.6\pm0.2$     & $1.51\pm0.15$   & $1.41\pm0.01$        & 1000         & $0.94\pm0.16$            & 0         &  \nodata               & 10,2,30       \\
2MASS J22393718+1617127	  &T3	        &$5.8\pm0.4$	&\nodata	    &\nodata	        &\nodata	        	&1000	         & \nodata           &0	        &\nodata	             &2              \\
2MASS J2244316+204343     & L6-L8   & $5.5\pm0.6$     & $0.8\pm$        &  \nodata          & $11\pm2$             & 120             & $1.28\pm0.02^8$           & 0         & $76^{+14}_{-20}$       & 31,8,5       \\
SDSSp J224953.45+004404.2 & L3+L5        &   \nodata       & \textless{}0.25 & \textless{}0.45 &  \nodata               & 1000       & \nodata             & 0         &  \nodata             & 2            \\
2MASSI J2254188+312349    & T5.0         &  \nodata      & \textless{}0.47 & \textless{}0.39 &   \nodata            & 1000           & \nodata             & 0         &   \nodata               & 2            \\
WISE J085510.83-071442.5  & Y2           &  \nodata         & $4\pm1$         & $4\pm1$         &   \nodata            & 1000        & \nodata              & 0         &  \nodata                & 32           \\
HD 203030B                & L7.5         & $1.1\pm0.3$     &   \nodata       &  \nodata          & $7.5\pm0.6$          & $9\pm60$   & \nodata             & 1         &  \nodata                 & 33      \\ 
\enddata
\tablecomments{Table available online. \\
a. Estimated radii are from \citet{Filippazzo2015}, except where a superscript indicates reference}
\tablerefs{
~(1)~\citet{Eriksson2019};
~(2)~\citet{Metchev2015a};
~(3)~\citet{Croll2016};
~(4)~\citet{Vos2017};
~(5)~\citet{Vos2019};
~(6)~This work;
~(7)~\citet{Lew2016};
~(8)~\citet{Vos2018};
~(9)~\citet{Naud2017};
~(10)~\citet{Radigan2014};
~(11)~\citet{Yang2016};
~(12)~\citet{Gagne2017};
~(13)~\citet{Apai2017};
~(14)~\citet{Buenzli2014};
~(15)~\citet{Clarke2008};
~(16)~\citet{Manjavacas2018};
~(17)~\citet{Gillon2013};
~(18)~\citet{Buenzli2015a};
~(19)~\citet{Girardin2013};
~(20)~\citet{Heinze2014};
~(21)~\citet{Schneider2018};
~(22)~\citet{Zhou2016};
~(23)~\citet{Bowler2020};
~(24)~\citet{Zhou2020};
~(25)~\citet{Manjavacas2019};
~(26)~\citet{Cushing2016};
~(27)~\citet{Leggett2016};
~(28)~\citet{Allers2016};
~(29)~\citet{Biller2018};
~(30)~\citet{Buenzli2012};
~(31)~\citet{Morales-Calderon2006};
~(32)~\citet{Esplin2016};
~(33)~\citet{Miles-Paez2019}}
\end{deluxetable}
\end{longrotatetable}
\clearpage

\clearpage
\appendix
\section{Posterior Distributions of Variability Parameters}

\begin{figure*}[b]
   \centering
   \includegraphics[scale=0.45]{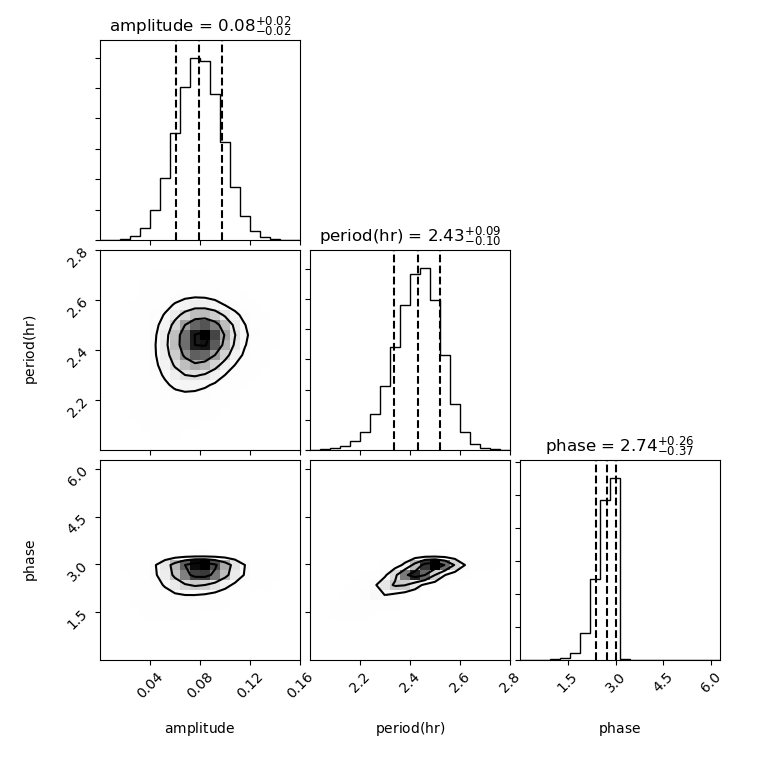}
    \includegraphics[scale=0.45]{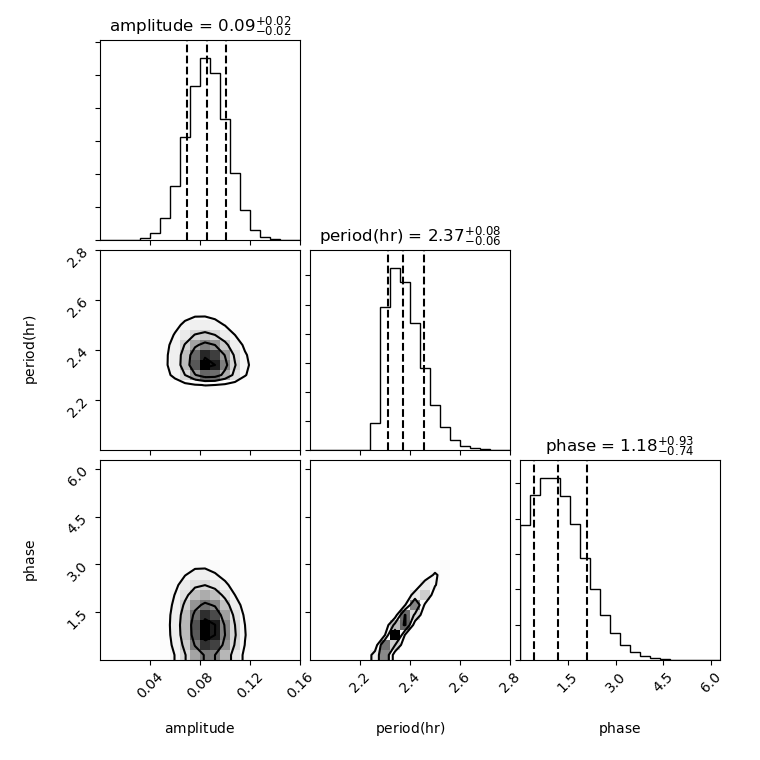}
   \caption{Posterior distribution of amplitude, period and phase parameters for \obj{2m0045}. Left panel shows parameters for Channel 2 data, right panel shows fit for Channel 1 data.}
   \label{fig:posterior_2M0045}
\end{figure*}

\begin{figure*}[b]
   \centering
   \includegraphics[scale=0.45]{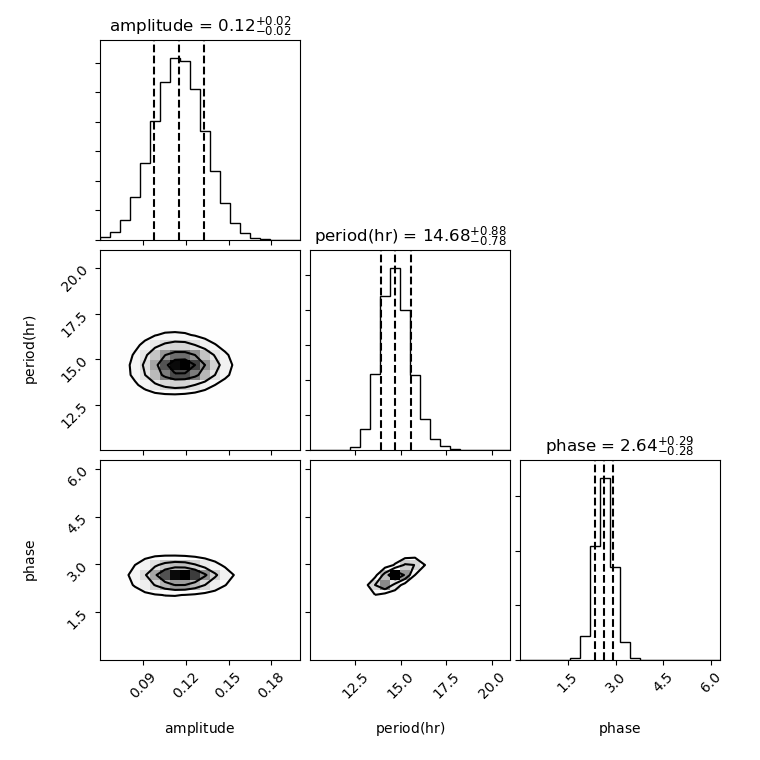}
    \includegraphics[scale=0.45]{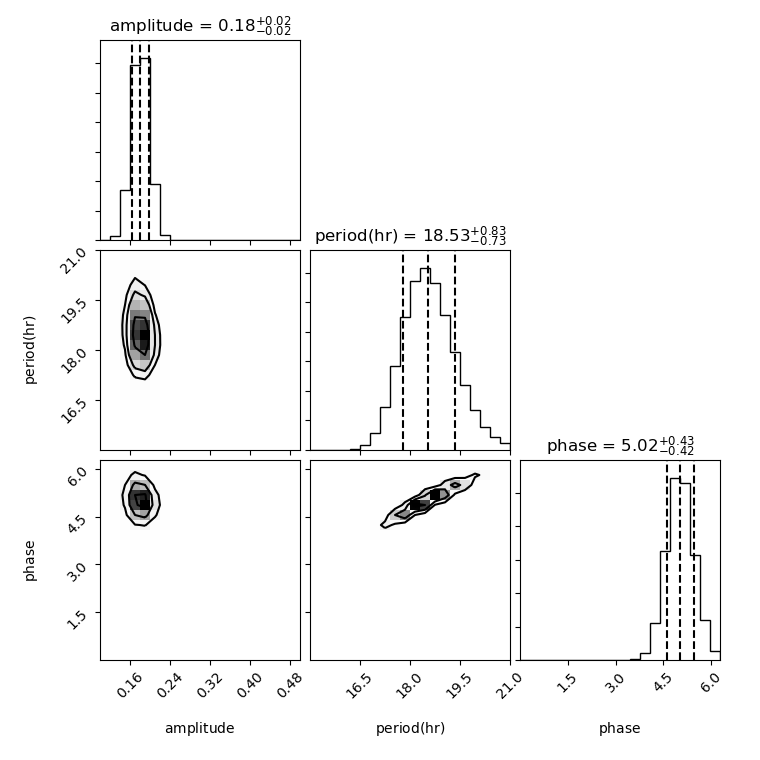}
   \caption{Posterior distribution of amplitude, period and phase parameters for \obj{2m0501}. Left panel shows parameters for Channel 2 data, right panel shows fit for Channel 1 data.}
   \label{fig:posterior_2M0501}
\end{figure*}

\begin{figure*}[tb]
   \centering
   \includegraphics[scale=0.45]{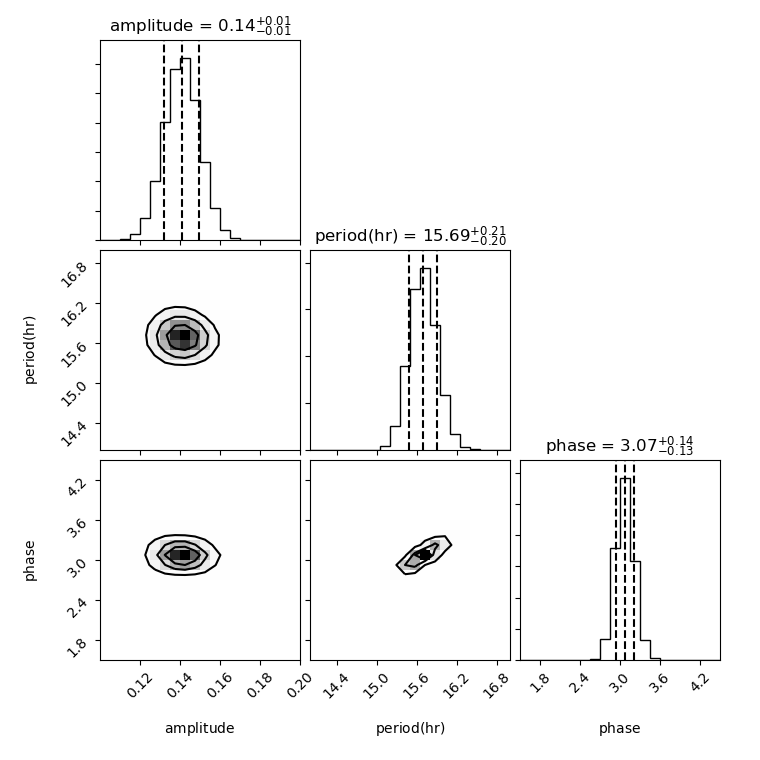}
   \caption{Posterior distribution of amplitude, period and phase parameters for sinusoidal fit to the full Channel 1 and Channel 2 light curve of \obj{2m0501}.}
   \label{fig:posterior_2M0501_full}
\end{figure*}

\clearpage
\section{Pixel Position Variations Over \textit{Spitzer} Variations}

\begin{figure*}[b]
   \centering
   \includegraphics[scale=0.9]{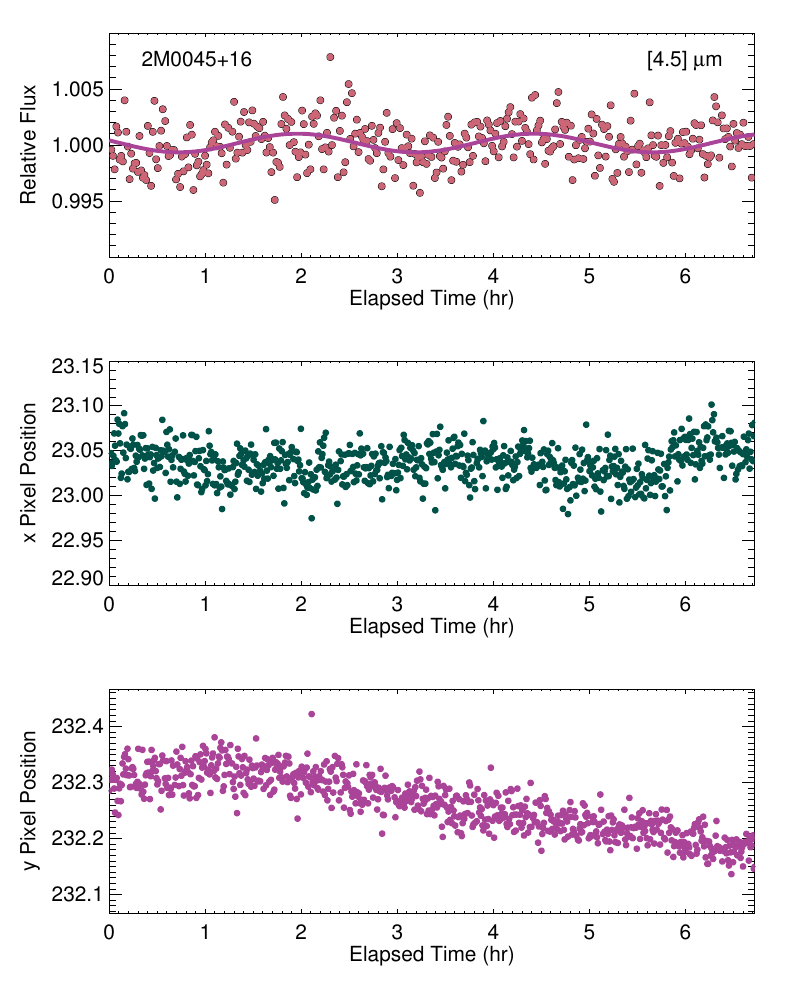}
    \includegraphics[scale=0.9]{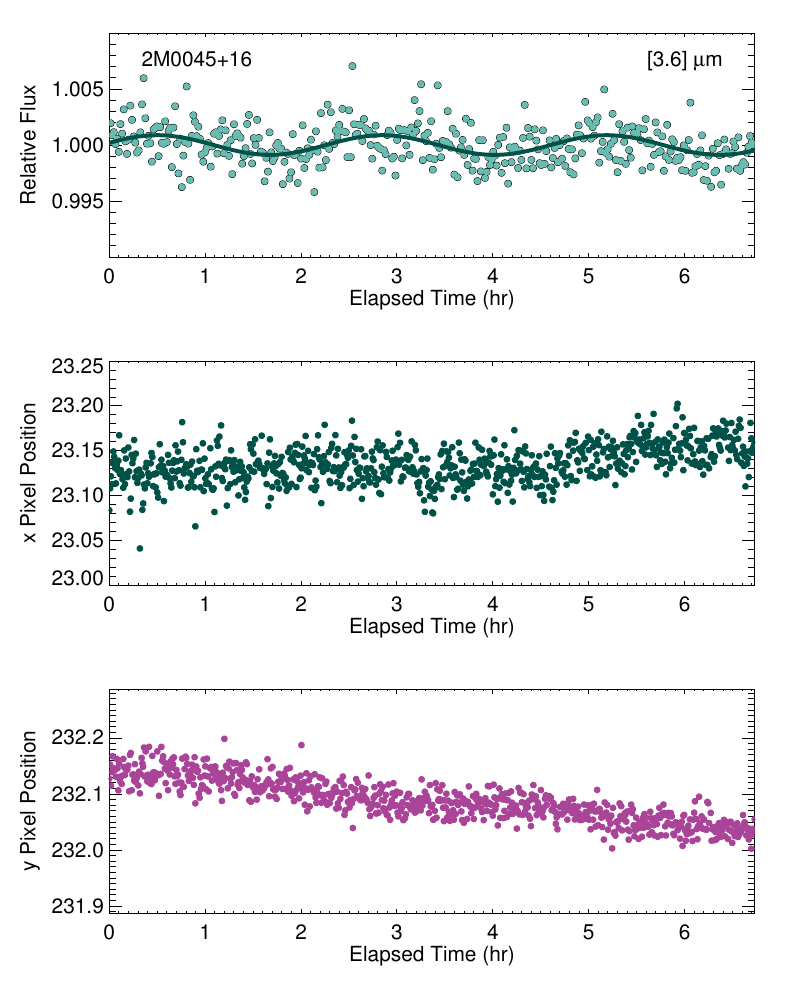}
   \caption{Corrected relative flux (top panel), x pixel position (middle panel) and y pixel position (bottom panel) for \textit{Spitzer} [$4.5~\mu$m] (left) and [$3.6~\mu$m] (right) monitoring of \obj{2m0045}. The observed variability is not correlated with the x,y pixel positions.}
   \label{fig:xypos_2M0045}
\end{figure*}

\begin{figure*}[b]
   \centering
   \includegraphics[scale=0.9]{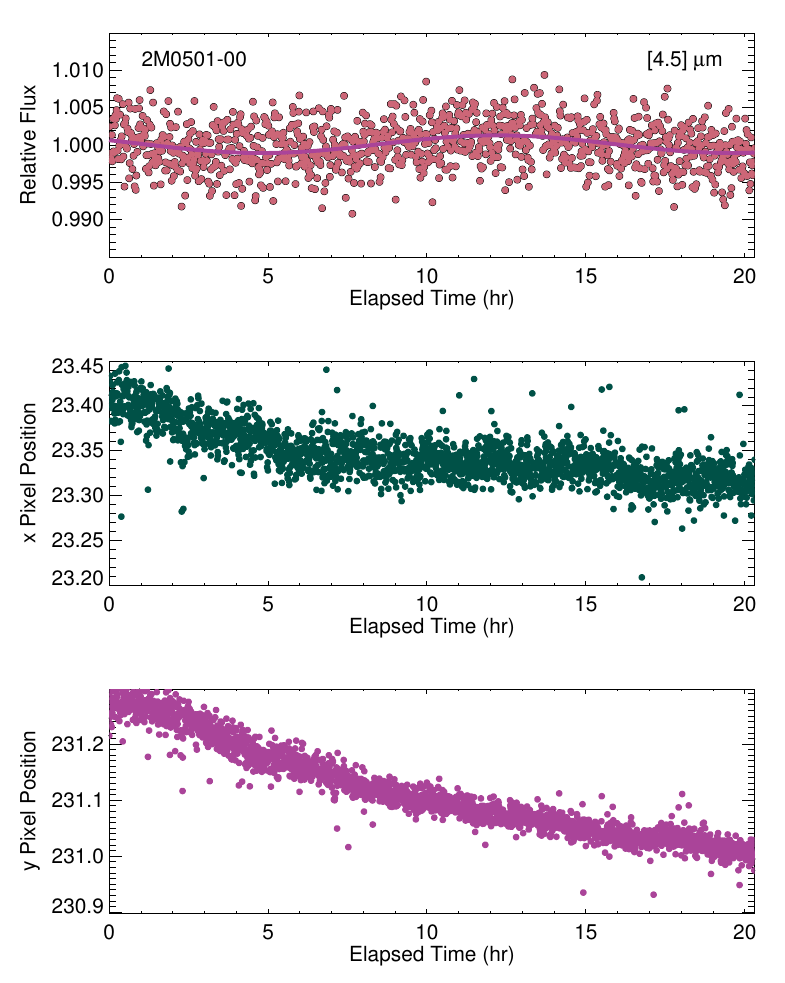}
    \includegraphics[scale=0.9]{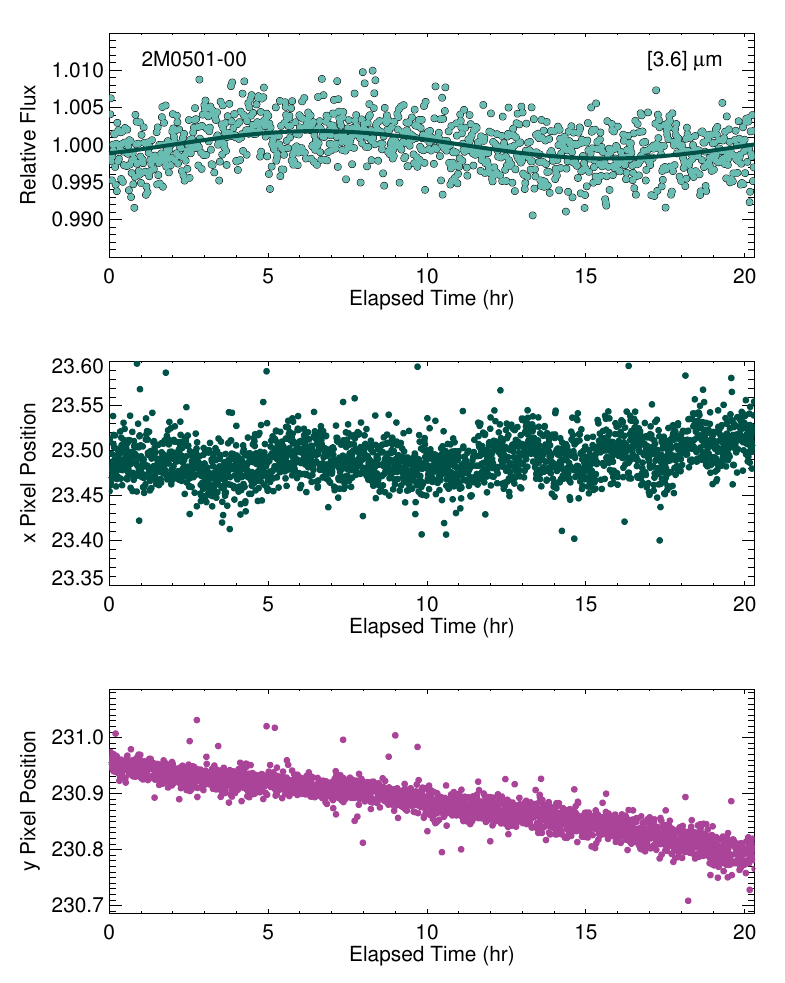}
   \caption{Corrected relative flux (top panel), x pixel position (middle panel) and y pixel position (bottom panel) for \textit{Spitzer} [$4.5~\mu$m] (left) and [$3.6~\mu$m] (right) monitoring of \obj{2m0501}. The observed variability is not correlated with the x,y pixel positions.}
   \label{fig:xypos_2M0501}
\end{figure*}

\begin{figure*}[tb]
   \centering
   \includegraphics[scale=1]{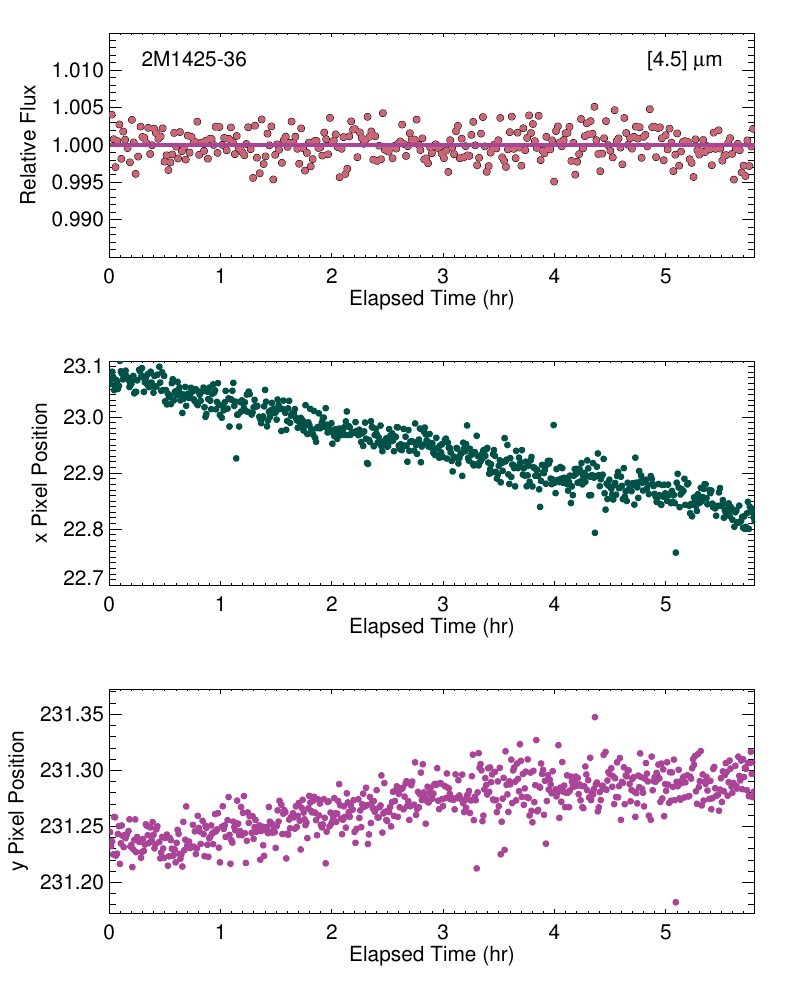}
    \includegraphics[scale=1]{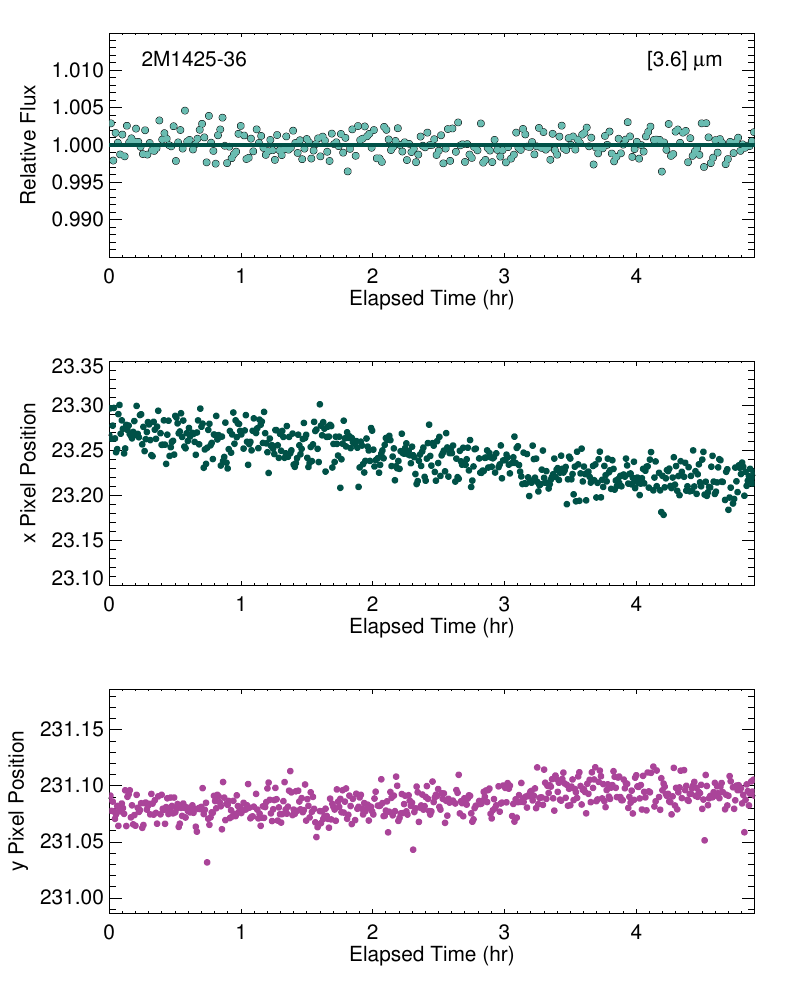}
   \caption{Corrected relative flux (top panel), x pixel position (middle panel) and y pixel position (bottom panel) for \textit{Spitzer} [$4.5~\mu$m] (left) and [$3.6~\mu$m] (right) monitoring of \obj{2m1425}. We do not detect significant variability is \obj{2m1425}.}
   \label{fig:xypos_2M1425}
\end{figure*}



\end{document}